\DocumentMetadata{}
\documentclass[nonacm,sigplan]{acmart}

\usepackage[]{hyperref}
\usepackage[normalem]{ulem}

\usepackage{caption}
\usepackage{graphicx}
\usepackage{textcomp}
\usepackage{xcolor}

\usepackage{listings}
\usepackage{booktabs}
\usepackage{multirow}
\usepackage{algorithm}
\usepackage{algpseudocode}
\makeatletter
\algrenewcommand\ALG@beginalgorithmic{\small}
\makeatother
\usepackage{enumitem}
\newcommand{\tool}{\textsc{GPUlog}}
\newcommand{\buffer}{Eager Buffer Management }

\newcommand{\yihao}[1]{{#1}}

\newcommand{\revised}[1]{{#1}}

\begin{document}

\title{Optimizing Datalog for the GPU}

% \author{...} % removed for anonymity
\author{Yihao Sun}
\email{ysun67@syr.edu}
\affiliation{%
  \institution{Syracuse University}
  \city{Syracuse}
  \state{New York}
  \country{USA}
}

\author{Ahmedur Rahman Shovon}
\email{ashov@uic.edu}
\affiliation{%
  \institution{University of Illinois, Chicago}
  \city{Chicago}
  \state{Illinois}
  \country{USA}
}

\author{Thomas Gilray}
\email{thomas.gilray@wsu.edu}
\affiliation{%
  \institution{Washington State University}
  \city{Pullman}
  \state{Washington}
  \country{USA}
}

\author{Sidharth Kumar}
\email{sidharth@uic.edu}
\affiliation{%
  \institution{University of Illinois, Chicago}
  \city{Chicago}
  \state{Illinois}
  \country{USA}
}

\author{Kristopher Micinski}
\email{kkmicins@syr.edu}
\affiliation{%
  \institution{Syracuse University}
  \city{Syracuse}
  \state{New York}
  \country{USA}
}

\begin{abstract}
 Modern Datalog engines (\emph{e}.\emph{g}., LogicBlox, Souffl\'e, \texttt{ddlog}) enable their users to write declarative queries which compute recursive deductions over extensional facts, leaving high-performance operationalization (query planning, semi-na\"ive evaluation, and parallelization) to the engine. Such engines form the backbone of modern high-throughput applications in static analysis, network monitoring, and social-media mining.
 In this paper, we present a methodology for implementing a modern in-memory Datalog engine on data center GPUs, allowing us to achieve significant (up to $45\times$) gains compared to Souffl\'e (a modern CPU-based engine) on context-sensitive points-to analysis of \textsf{httpd}.
 We present \tool{}, a Datalog engine backend that implements iterated relational algebra kernels over a novel range-indexed data structure we call the hash-indexed sorted array (HISA). HISA combines the algorithmic benefits of incremental range-indexed relations with the raw computation throughput of operations over dense data structures. Our experiments show that \tool{} is significantly faster than CPU-based Datalog engines while achieving a favorable memory footprint compared to contemporary GPU-based joins.
\end{abstract}

\maketitle % should come after the abstract
\pagestyle{plain} % should come right after \maketitle

\section{Introduction}
\label{sec:intro}

Declarative languages enable their users to write high-level logical rules that specify acceptable solutions to a problem, leaving efficient implementation to a high-performance backend. In particular, modern CPU-based Datalog engines power state-of-the-art systems in program analysis~\cite{bravenboer2009doop,flores2020ddisasm,balatsouras2016cclyzer,recstep}, social-media mining~\cite{shkapsky2016big,gu2019rasql,socialite, skvortsov2024logica}, and  business analytics~\cite{business-Datalog}. Such engines enable a user to specify deductions (forming an \emph{intensional} database, henceforth IDB) over extensionally manifest relations (the \emph{extensional} database, henceforth EDB). For example, given an input relation \textit{Edge}(\texttt{from},\texttt{to}), the program \textit{REACH} computes
\textit{Edge}'s transitive closure:
\[
\begin{array}{lcl}
    \textit{Reach}(\texttt{from},\texttt{to})  & \!\!\!\! \leftarrow & \!\!  \textit{Edge}(\texttt{from}, \texttt{to}).   \\
    \textit{Reach}(\texttt{from}, \texttt{to}) & \!\!\!\! \leftarrow & \!\! \textit{Edge}(\texttt{from}, \texttt{mid}), \textit{Reach}(\texttt{mid}, \texttt{to}).
\end{array}
\]

The second rule requires recursive computation to a fixed-point: the engine iteratively discovers an ever-larger intensional database (\textit{Reach}), starting from the extensional database (\textit{Edge}).
Unfortunately, best-in-class in-memory Datalogs hit scalability walls around $8$--$16$ threads due to the challenges of working with locking, linked data structures in a parallel, shared-memory setting. For example, as we will soon see, when run at $32$ threads on transitive closure, Souffl\'e \cite{jordan2016souffle} (a state-of-the-art CPU-based engine) spends $77.8$\% of its time in serialized tuple deduplication/insertion. 

Compared to CPUs, GPUs offer hundreds of thousands of threads, along with extremely high memory throughput via HBM~\cite{hbm}. However, achieving optimal performance necessitates embracing the GPU's programming model. Traditionally, CPU-based graphics processing code has been accelerated via SIMD instructions (such as AVX and SSE), necessitating that all threads operate in lockstep and strongly penalizing thread divergence. While GPUs are inspired by this SIMD paradigm, modern GPGPU programming is SIMT in nature, allowing parallel execution at a per-thread granularity. Threads can diverge at the sub-warp level using advanced thread scheduling techniques~\cite{simt}.
As we will see, the SIMT nature of modern GPUs is a natural fit for Datalog, which stresses both massive data parallelism (\emph{e}.\emph{g}., large joins), \emph{and} task parallelism (\emph{e}.\emph{g}., multiple rules).

% By contrast, GPUs have \yihao{hundreds} thousands of threads and AL; clearly, CPU-based approaches will not suffice to enable scalable Datalog on the GPU.
% The crux of the matter is that Datalog's dense loop is range-indexed joins: unlike in SQL (where indices may not be known a-priori), Datalog engines ensure each relation is indexed so that it may always be efficiently range-queried rather than iterated.

% Modern GPUs provide thousands of parallel data threads in their high-throughput SIMD design, making them well-suited for implementing high-performance Datalog engines. Implementing Datalog requires executing key relational algebra operations like join, union, and projection, along with tasks like sorting and merging buffers \emph{iteratively}. Most prior work has focused on optimizing these tasks in isolation via batch processing. Performing these tasks iteratively and scalably while accounting for the SIMD architecture and restrictive memory patterns of GPUs presents challenges. In this work, we introduce \tool{}, a GPU-accelerated Datalog engine that achieves roughly 10x runtime speedups versus an optimally configured state-of-the-art CPU engine (Souffl\'e) on large deductive-analytic workloads (\emph{e}.\emph{g}., context-sensitive points-to analysis of \texttt{httpd}, see Section~\ref{sec:eval} for our results).

In this paper, we introduce \tool{}: a GPU-based library for parallel relational algebra kernels, enabling the execution of Datalog programs on modern GPUs. \tool{} is backed by a novel SIMT data structure: the hash-indexed sorted array (HISA). We designed HISA to balance the concerns of modern Datalog on the GPU, enabling three critical tasks: (1) efficient range queries, necessary for joins, (2) lock-free deduplication, and (3) parallel iteration. Additionally, \tool{} provides abstractions for optimized evaluation of $n$-way loop-joins we call \emph{temporarily-materialized} joins, offering a space-for-time trade-off we found crucial to scale to high data loads.

We have used \tool{} to implement and comprehensively study the performance of graph analytics (reachability, same generation) and program analysis (points-to analysis), evaluating \tool{} against cuDF, GPUJoin, and Souffl\'e \cite{cudf, shovon2023towards, jordan2016souffle}. We observe improved ($5\times$) join performance compared to state-of-the-art GPU join algorithms (due to efficient range queries enabled by HISA), along with significantly reduced memory footprint. Additionally, \tool{} is the first-ever GPU Datalog implementation to achieve net-positive performance versus CPU-based engines, beating Souffl\'e (a leading CPU-based solver) by up to $45\times$ (NVIDIA H100 \tool{} vs. EPYC 7543P Souffl\'e). Our contributions are as follows:
\begin{itemize}
%\item A systematization of the key differences and necessities of in-memory Datalog engines in the SIMD/GPU vs. shared-memory/CPU setting.
\item The Hash-Indexed Sorted Array (HISA), a novel relation-backing data structure which provides range querying while leveraging the efficiencies of dense representations.

\item \tool{}, a CUDA-based library for implementing Datalog queries on the GPU; \tool{} uses HISA as its tuple representation. \tool{} also leverages two novel strategies apropos Datalog on the GPU: eager buffer management and temporarily-materialized $n$-way joins. 

\item A thorough evaluation of \tool{}, and a performance comparison between \tool{} and state-of-the-art CPU and GPU-based Datalog engines and GPU joins; we show \tool{} outperforms prior work by $5$--$45\times$ with favorable memory footprint.

% \item An evaluation of \tool{} which evaluates its performance as a high-throughput SIMD hash table.

% \item An evaluation of our eager buffer management strategy, showing that it significantly improves buffer allocation.

% \item An evaluation comparing \tool{} to both CPU and GPU-based systems for deductive-analytic queries, showing that \tool{} outperforms every system against we evaluate (often by a significant factor). 
%Next , we discuss how we use HISA to implement \tool{}, 

\end{itemize}

%We begin by surveying the design and implementation of modern CPU-based deductive database systems, particularly remarking upon their scalability 
%bottlenecks. We next shift to discussing concerns regarding a relation-backing SIMD data structure for iterated joins on the GPU, before introducing our solution (the HISA) in Section~\ref{sec:impl-data}. Section~\ref{sec:gdlog} then discusses how we leverage HISA to build \tool{}. Our evaluation is in Section~\ref{sec:eval}, after which we conclude.

\section{Datalog and Declarative Analytics}
\label{sec:datalog}

Datalog has yielded exciting results for a diverse set of data-heavy analytic applications such as static program analysis~\cite{bravenboer2009strictly}, graph mining~\cite{seo2013socialite}, and machine learning~\cite{
mooney1996inductive}. Its growing popularity lies in its expressive and elegant semantics for specifying computations. Datalog programs comprise an extensional database (EDB) of explicit facts and an intensional database (IDB) of derived facts, transitively inferred from the rules and EDB~\cite{green2013datalog}. The language's rules, written as Horn clauses, define relationships between data, and Datalog operates by iteratively applying these rules until no more knowledge is discovered~\cite{maier2018datalog}. Each rule consists of a head and a body. The general form of a Datalog rule is:
\[
\begin{array}{lcl}
 \textit{Head}(...)  & \leftarrow & \textit{Body}_1(...), ..., \textit{Body}_n(...).
\end{array}
\]
The head, represented by a single predicate atom, signifies the derived fact to be inferred. The body, made up of comma-separated predicate atoms, specifies the conditions for the head's truth, with commas serving as a logical "AND" ($\land$). 
The implication symbol $\leftarrow$ connects the head and body, signaling that the head is derived when the body's conditions are met. This structure is closely related to concepts in relational database management systems (RDBMS). Notably, the sharing of logical variables across body clauses directly corresponds to a \emph{join} operation on the related columns in an RDBMS. However, Datalog's foundation in logic programming transcends the capabilities of traditional SQL. It leverages this foundation to unlock greater expressive power, enabling functionalities like recursion and inference beyond simple data retrieval~\cite{10.1145/2892208.2892226}.

% For instance, consider the second rule of the REACH query introduced in beginning of section \ref{sec:intro}. Here, the ``mid'' variable, shared between clauses, indicates a join between the second column of the \textit{Edge} table and the first column of the \textit{Reach} table. The results of this join, following the logic of all body clauses, are then inserted into the table associated with the head clause. 

\paragraph*{Semi-na\"ive evaluation}
Modern Datalog engines owe their algorithmic benefits to incremental evaluation techniques such as semi-na\"ive evaluation~\cite{abiteboul1995foundations}, differential/timely dataflow~\cite{Naiad, mcsherry2013differential}, and DBSP~\cite{dbsp}. Following Souffl\'e, \tool{} uses semi-na\"ive evaluation, which builds a \emph{frontier} of freshly-discovered facts, avoiding the inevitable re-discovery of a fact at every subsequent iteration. To illustrate this process, consider the Same Generation (\emph{SG}) query, used to determine if two nodes in a graph share a topological order:
\[
\begin{array}{lcl}
    \textit{SG}(\texttt{x},\texttt{y}) & \leftarrow & \textit{Edge}(\texttt{p},\texttt{x}), ~ \textit{Edge}(\texttt{p}, \texttt{y}), ~ \texttt{x} \neq \texttt{y}. \\
    \textit{SG}(\texttt{x}, \texttt{y}) & \leftarrow & \textit{Edge}(\texttt{a}, \texttt{x}), ~ \textit{SG}(\texttt{a},\texttt{b}), ~ \textit{Edge}(\texttt{b},\texttt{y}), \texttt{x} \neq \texttt{y}.
\end{array}
\]
% Here, the \textit{Edge} relation stores all edges within the input graph. When nodes $x$ and $y$ are determined to be of the same generation (implying a shared topological ordering) an \textit{Edge} $(x, y)$ fact will be established in the database. 
In this query, \textit{Edge} represents extensional edges in the input graph. The first rule states that nodes \texttt{x} and \texttt{y} are in the same generation if they have a common parent node \texttt{p}. The second rule recursively derives \textit{SG} by determining that \texttt{x} and \texttt{y} are in the same generation if there exists a node \texttt{a} with an outgoing edge to \texttt{x}, and another node \texttt{b} with an outgoing edge to \texttt{y}, such that \texttt{a} and \texttt{b} are themselves in the same generation.

The semi-na\"ive evaluation strategy optimizes this query execution, by maintaining the \textit{SG} relation in three versions: ``\textit{new}'' (containing tuples generated in the current iteration), ``\textit{delta}'' (holding unique tuples added in the previous iteration), and ``\textit{full}'' (comprising all tuples derived across all iterations). By performing the join operation solely on the \textit{delta} relation, the engine significantly reduces redundant computation.
% leading to substantial performance gains over na\"ive evaluation strategies.
% We visualize this strategy by Figure~\ref{fig:sg-iter1} and Figure~\ref{fig:sg-iter2}.

\begin{figure}
    \centering
    \includegraphics[width=\linewidth]{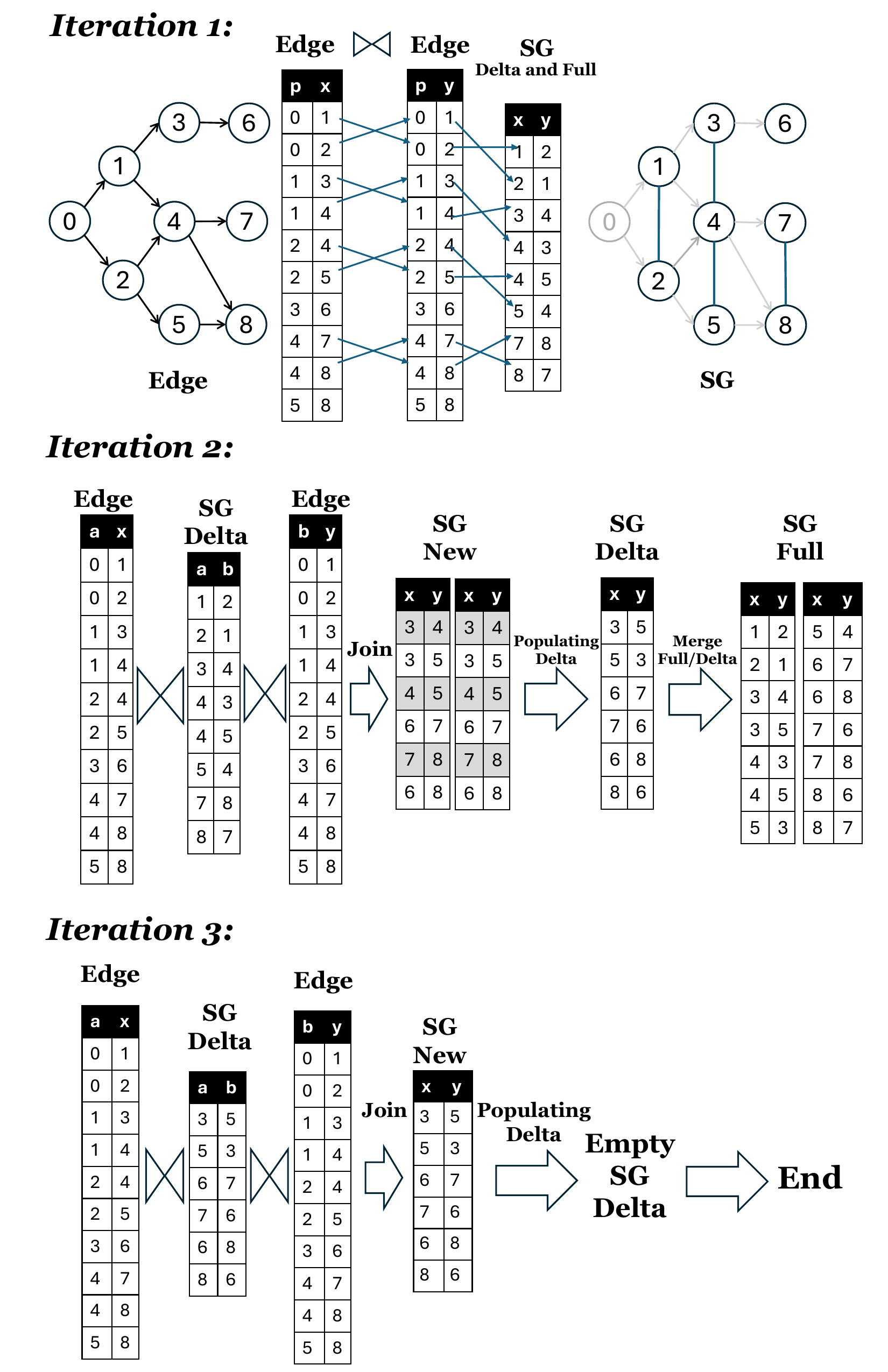}
    \caption{\fontsize{9}{11}\selectfont The execution results of each iteration in Same Generation (SG) query.}
    \label{fig:sg-iter}
    % \vspace{-0.5cm}
\end{figure}

\paragraph*{Example} Figure~\ref{fig:sg-iter} illustrates each iteration of the \textit{SG} query. During the first iteration, only the first rule generates new tuples since \textit{SG} is empty and thus the second rule yields nothing.
This rule applies when two edges in the graph merge from the same starting node but lead to different destination nodes; these edges are joined to generate new \textit{SG} tuples. For instance, the tuple $\textit{SG}(7,8)$ is produced by joining $\textit{Edge}(4,7)$ and $\textit{Edge}(4,8)$, both of which originate from node $4$. Upon completing the first iteration, the newly generated tuples are moved into both the \textit{delta} and \textit{full} versions of the \texttt{SG} relation, preparing them for subsequent computation.

% \begin{figure}
%     \centering
%     \includegraphics[width=\linewidth]{figures/sg_iter2.pdf}
%     \caption{The join and deduplication operation in the $2^{nd}$ of the Same Generation Query. The grey rows in $\textit{SG}_{new}$ are duplicates.}
%     \label{fig:sg-iter2}
% \end{figure}
%Since, at the end of the first iteration, the relation $SG$ gets populated delta where the SG relation's inductive cases are engaged.
The middle of Figure~\ref{fig:sg-iter} shows the second iteration involving a three-way join: $\textit{Edge} \bowtie \textit{SG}_{delta} \bowtie \textit{Edge}$, leading to the derivation of indirect same generation tuples, such as $\textit{SG}(6,8)$, from $\textit{Edge}(3,6)$, $\textit{Edge}(4,8)$, and $\textit{SG}(3,4)$ (in the \textit{delta} version). Notice that the \textit{new} version of the \texttt{SG} relation, generated post-join, contains duplicates tuples from its \textit{full} version, such as $\textit{SG}(3,5)$ and $\textit{SG}(7,8)$.
To ensure maintain the invariant that \textit{delta} and \textit{full} are disjoint, a deduplication process is implemented both within the \textit{delta} and against the \textit{full} set as the \textit{new} tuples are allocated.

% \begin{figure}
%     \centering
%     \includegraphics[width=\linewidth]{figures/sg_iter3.pdf}
%     \caption{The join and deduplication operation in the $2^{nd}$ of the Same Generation Query.}
%     \label{fig:sg-iter3}
% \end{figure}

% \kris{yihao: edit plz, then remove this comment.}
After completing the second iteration, the Datalog engine proceeds with the third iteration, as shown in the bottom of Figure~\ref{fig:sg-iter}. The previous iteration's $\textit{SG}_\textit{delta}$ is used as the input for the join operations in the current iteration. The join results in new tuples $\textit{SG}_{new}$; however, all of these tuples are already present in $\textit{SG}_{full}$, leading to an empty $\textit{SG}_{delta}$. This indicates that the query has reached its fixpoint.  We revisit semi-na\"ive evaluation in more detail within the context of our GPU-based implementation in Section~\ref{sec:impl-data}.

\section{GPU Datalog: Challenges and Concerns}
\label{sec:motivation}

%\kris{inner and outer is no longer defined}

The high performance of modern Datalog engines (such as Souffl\'e~\cite{jordan2016souffle} and BigDatalog~\cite{shkapsky2016big}) is due to a combination of factors, including compilation, data structures, and semi-na\"ive evaluation. Among these, we focus on the design of a GPU-based relation-backing data structure. In the context of relevant related work, we outline four key requirements demanded of a GPU-based relation-backing data structure. No state-of-the-art implementations sufficiently satisfy all of our requirements (at least, in a fully-general manner).

\paragraph*{[R1]: Efficient Range-Querying} Joins $R(\ldots) \bowtie Q(\ldots)$ are  operationalized in modern Datalog engines via loop-joins, scanning $R$ and range-querying $Q$, or mutatis mutandis for $Q$ and $R$. 
To support range-querying, tuples are organized (\emph{e}.\emph{g}., via indexing) to associate sets of tuples with a set of join columns. We call the scanned relation the \emph{outer} relation, and the range-indexed relation the \emph{inner} relation. Compared to traditional RDBMS systems---which cannot possibly compute optimal indices due to the ad-hoc nature of querying---modern Datalog engines perform indexing for \emph{every} query~\cite{indexselection}.
Modern engines employ a mix of minimally-locking linked data structures such as B-trees and Bries \cite{jordan2019specialized,jordan2019brie,byods}, and are carefully tuned for optimal performance on shared-memory, CPU-based systems.

We can now identify the first requirement for our GPU-based data structure: it must \emph{support fast range queries [R1]}. Several prior GPU-based approaches (\emph{e}.\emph{g}., GPUJoin) have implemented SIMT-friendly data structures to enable fast range queries. For example, relations may be stored in hashmaps whose indexed columns serve as keys, and whose non-indexed columns act as values. By leveraging atomic operations for key updating and open addressing, these hashmaps may be constructed in a data-parallel manner.

\paragraph*{[R2]: Parallel Iteration} Hash-based data structures, due to their sparse nature, face limitations in efficient iteration. To enable parallel iteration on the GPU, the data stored in these hash-based structures must undergo a serialization process. This involves traversing the hashmaps and copying all the elements into a contiguous array. Consequently, while these data structures are suitable for inner relations where range query speed is the focus, they are not ideal for outer relations that require iteration. It can be argued that keeping the outer relation as an array and the inner as a hash table could resolve this issue. However, in Datalog, it is common for a relation to function both as an inner and outer relation. 
As an example, the first rule of the same generation query in Section~\ref{sec:datalog}, involves a self-join on the Edge relation.

%For instance, the first rule of the SG relation is discussed in Section~\ref{sec:datalog}, where the \texttt{Edge} relation joins with itself.

% The limitations of these hash-based data structures, particularly their iterability, make them less favorable for serving as outer relations in join operations, where GPU-parallelizable traversal of the data is crucial for optimal performance. 
% And, this brings us to our second requirement for the data structure, it should \emph{support parallel iteration [R2]}.
The limitations of hash-based data structures, particularly their iterability, make them less favorable for outer relations in join operations where GPU-parallelizable traversal is crucial. Thus, our second requirement for the data structure is to support parallel iteration [R2].

%Using hashmaps in GPU-based Datalog engines enhances indexing speed for parallel joins, but poses challenges as the outer relation due to the SIMD nature of GPUs. In systems like GPUJoin, placing an indexed relation outside requires transforming it into a contiguous array for optimal GPU performance, as GPUs operate best with sequentially arranged data. This extra need for serialization can degrade join performance and restrict the engine's capacity for real-world queries. 

%This underscores the need to investigate alternative or hybrid data structures that align with GPU processing strengths, while accommodating the irregular access patterns inherent in Datalog queries more complex than reachability analysis.

\paragraph*{[R3]: Multiple Join Columns} 
%The atomic variable's size cannot increase to accommodate the arbitrary size of joined columns, posing a challenge for developing a GPU-based Datalog engine capable of handling common 
Real-world analysis queries can contain multiple join columns. For instance, consider the query from DDisasm \cite{flores2020ddisasm} (a Datalog-based disassembler):
\[
\begin{array}{ll}
\textit{value\_reg\_unsupported} ( & \! \! \! \! \! \! \! \! \! \! \! \!  \texttt{EA},\texttt{Reg}) \quad  \leftarrow  \\
\qquad \textit{def\_used.for\_address} ( & \! \! \! \! \! \! \texttt{EA},\texttt{Reg},\_), \\
\qquad \textit{arch.memory\_access} ( & \! \! \! \! \! \!  \textsc{LOAD},\texttt{EA}, \\
   & \! \! \! \! \! \_,\_,\texttt{Reg},\texttt{RegBase},\_,\_,\_), \\
\qquad \textit{RegBase} \neq \textsc{NONE} & \! \! \! \! \! \! \!.
\end{array}
\]
This \emph{join} operation involving the  \textit{def\_used.for\_address}
and \textit{arch.memory\_access} is executed on two columns: \texttt{EA} and \texttt{Reg}. Parallel construction of hashmaps on GPUs depends on atomic operations, which have a size limitation of $64$ bits, or at most, $128$ bits. As a result, directly using multiple columns as keys in the hashmap can be problematic, especially when the combined size of these columns exceeds the $128$-bit limit. This restriction poses challenges in efficiently storing and accessing data with multiple columns in GPU-based hashmap implementations. This brings us to the third requirement of our data structure, it must \emph{support multiple join columns [R3]}.

%as join columns at the same time

%the relation \textit{value\_reg\_unsupported} captures instances where a value is generated for a destination register \texttt{Reg} at a code address \texttt{EA} through a load instruction from another register, and not from a constant address. This scenario is flagged as an unsupported indirect value flow, which will be treat as leaf node in DDisasm's value flow graph. The \textit{arch.memory\_access} relation in DDisasm, utilizing both \texttt{EA} and \texttt{Reg} as join columns at the same time, presents a challenge for the relation storage in previous fast GPU join.

\paragraph*{[R4]: Efficient deduplication}
 As a final requirement, the data structure should \emph{support efficient deduplication}~[R4]. As mentioned earlier, deduplication is a crucial part of semi-na\"ive evaluation in modern Datalog engines (see Figure~\ref{fig:sg-iter}). Unfortunately, many existing GPU Datalogs do not support deduplication (RedFox~\cite{wu2014redfox} and GPUDatalog~\cite{martinez2013Datalog}).

\section{The Hash-Indexed Sorted Array (HISA)}
\label{sec:impl-data}
% \label{sec:impl}

\begin{figure}[t]
    \centering
    \includegraphics[width=\linewidth]{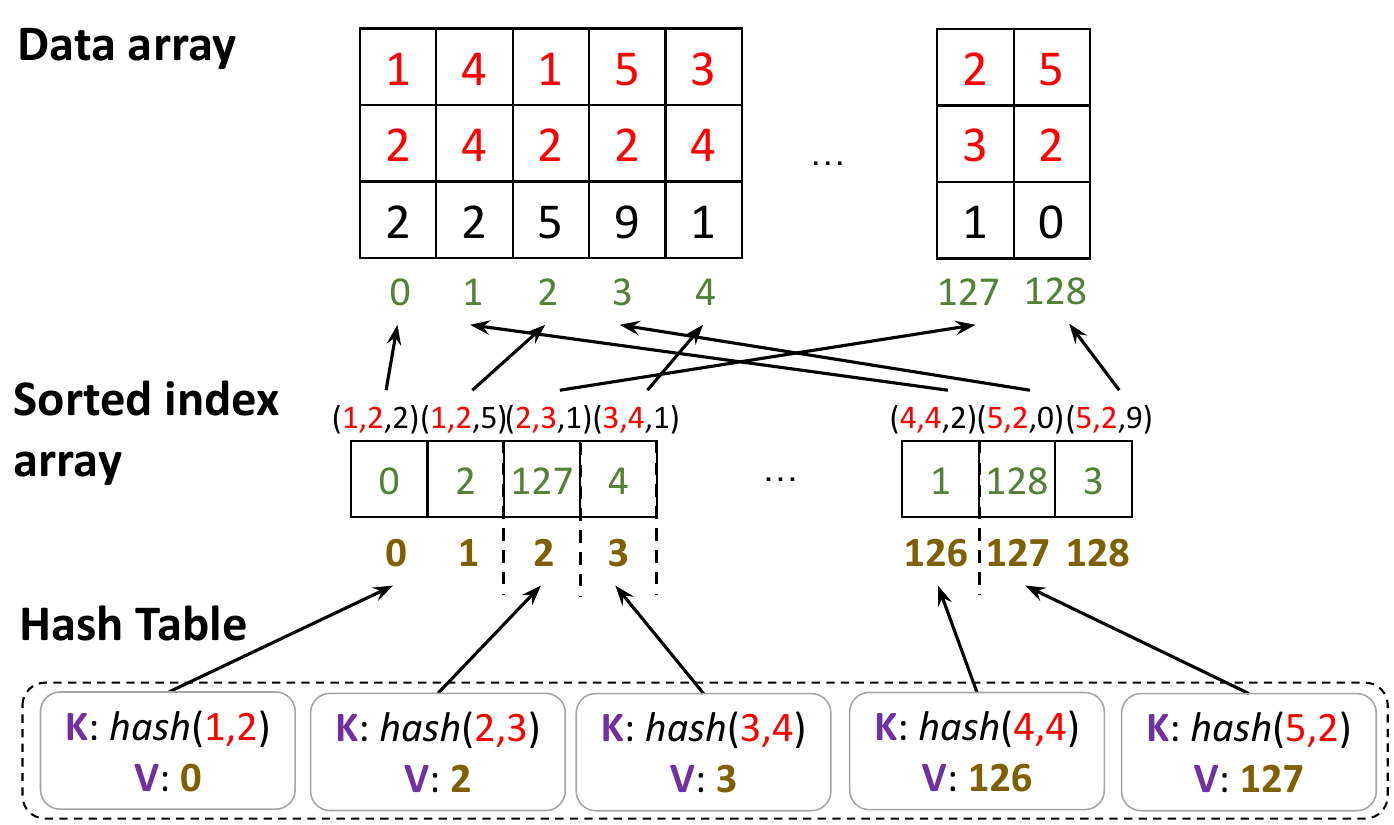}
    \caption{\fontsize{9}{11}\selectfont Example of a 3-arity relation with 2 join-columns (shown in color red) stored in HISA.}
    \label{fig:hisa}
    % \vspace{-0.5cm}
\end{figure}

To meet all four requirements ([R1], [R2], [R3], and [R4]), we developed a novel data structure we call the Hash-Indexed Sorted Array (HISA).
% \subsection{Key components of HISA}
%HISA is a three-layered data structure, comprising a data array, that stores the original tuple data, a sorted index array that stores the indices of the tuples in increasing order, and an open-addressing hash table, built on top of the sorted index array.
%Unlike other data structures where the hash map is built on top of the actual data (tuples), we build our hash map on the sorted index array, while only storing the reference for tuples with unique join column keys. This novel design of the data structure enables fast iteration and fast-range queries while also facilitating deduplication.
%HISA's novel design lies in the way the hash map is built on top of the sorted index array, while only storing the first of the 
%The data array enables fast iterations, the sorted index array is crucial for deduplication, and the hash map enables fast-range queries and facilitates joins on multiple columns. 
HISA, \revised{inspired by~\textit{HashGraph}~\cite{10.1145/3460872}}, is a custom-designed data structure consisting of three interconnected layers. At its foundation lies the data array, which stores the original tuple data. Built upon this data array is a sorted index array, which stores the indices of the tuples in ascending order.
%The tuple indices in the sorted index array are arranged in ascending order based on a lexicographical ordering of the tuples, where the columns are reordered such that the join columns appear first, followed by the remaining columns. 
The third layer is an open-addressing hash table, constructed on top of the sorted index array. HISA is different from other data structures as instead of building the hash table directly on the actual data (tuples), we construct it on the sorted index array. This approach allows us to store references only for tuples with unique join column keys, resulting in a more efficient use of memory.

% Secondly, it supports range-queries, as the hash table built on the sorted index array allows for quick lookups of specific subsets of data. Lastly, by storing the tuples in a sorted order, HISA inherently facilitates deduplication, eliminating redundant data and optimizing storage efficiency.

Figure~\ref{fig:hisa} sketches an example of a HISA data structure at runtime (in VRAM). We will now walk through each of the tiers, and discuss how they collectively enable [R1]--[R4]. 

%Figure~\ref{fig:hisa} shows the key components of HISA. We now present details of each of these components.
%Deduction-heavy workloads on GPUs demand efficient data structures that enable indexing, sorting, and range queries. To meet these requirement, we introduce the Hash-Indexed Sorted Array (HISA), a hybrid SIMD data structure designed for optimal execution of deduction-heavy workloads on the GPU.
%It is inspired by the compact and serializable nature of \textit{HashGraph}~\cite{10.1145/3460872} while tailoring it for Datalog evaluation. At a high level, HISA combines a data array and a sorted index array to support parallel iteration and deduplication, alongside an open-addressing hash table to enable efficient k-arity joins across multiple columns while also facilitating deduplication.

%HISA achieves its performance optimization through a meticulously crafted integration of algorithmic components and data structures.

% a novel k-arity data structure designed for efficient k-arity joins with multiple join columns on the GPU. This data structure combines the benefits of hash tables and sorted arrays which are well-suited for exploiting data parallelism utilizing the GPU architecture. The design of the data structure has three primary steps: 

\subsection{Data Array} 
The \emph{data array} represents the original GPU buffer that stores all the tuples of a relation transferred from the CPU to the GPU.
%The \emph{data array} corresponds to the original GPU buffer that holds all the tuples of a relation passed to the GPU from the CPU.
The $k$-ary relation data can be seen as a 2D array of size $n \times k$, where $n$ is the total number of tuples/rows and $k$ is the number of columns of the relation. An example of this 2D array can be seen at the top of Figure~\ref{fig:hisa}. Here $n = 129$ and $k = 3$, but, for the sake of clarity in exposition, the 2D array is shown in its transposed view. We store this 2D data in row-major order in the GPU buffer. Therefore, the tuples of Figure~\ref{fig:hisa} are stored in a buffer that looks like -- $\{1, 2, 2, 4, 4, 2, 1, 2, 5 \dots, 2, 3, 1, 5, 2, 0\}$.

This densely-packed contiguous arrangement, in contrast to sparse data structures, simplifies access for the GPU. GPU threads can then efficiently perform strided access to underlying the data array.

This simple layout makes it easy to implement parallel data retrieval, addressing the requirement for fast iteration [R2]. 
Coalesced memory access, enabled by our layout, allows multiple GPU threads to fetch data from the same memory block simultaneously, improving cache performance and optimizing memory operations. We provide implementation details of this parallel data access in Section~\ref{sec:gdlog-eval}, in the context of iterating the outer relation for a join operation.

\paragraph*{Merge operation}
\revised{Merging two HISAs typically occurs when combining the full and delta versions of a relation. In semi-na\"ve evaluation, to avoid redundant computation, the delta relation contains only fresh and unique tuples generated in the previous iteration, deduplicated against the full relation. Consequently, when merging delta and full, no additional deduplication is needed—the delta’s data array can be directly concatenated to the full’s data array.}

\begin{algorithm}[t]
\caption{Construction of Sorted Index Array $H_i$ for Relation $H$}
\label{alg:index-construct}
\begin{algorithmic}[1]
% \State Create a new HISA object $H_{i}$
\For{ all tuple $T$ in $H.\text{data\_array}$ \textbf{parallel} }
    \State $\textit{jc} \gets $ [join columns of $T$]
    \State $\textit{njc} \gets $ [non-join columns of $T$]
    \State Insert ($\textit{jc}$ ++ $\textit{njc}$) into $H_{i}.\text{data\_array}$
\EndFor
\State $H_{i}.\text{sorted\_index\_array} \gets $ [0 ... $H.\text{tuple\_size}$]
\For{$i$ in [0 ... $H.\text{arity}$]}
    \State $col_{tmp} \gets$ the $i_{th}$ of column of tuples in $H_i$
    \State stable sort $H_{i}.\text{sorted\_index\_array}$ use $col_{tmp}$ as key
\EndFor
\end{algorithmic}
\end{algorithm}
%\vspace{-0.3cm}

\subsection{Sorted Index Array}

The sorted index array stores the indices of the tuples from the data array in ascending order. The indices are arranged based on a lexicographical ordering of the tuples. This ensures the join-columns appear first, followed by the remaining columns. For example, consider a $3$-arity data array with tuples $\{2,1,5\}$, $\{2,5,9\}$, and $\{2,1,2\}$, where the second column is the join-column. The corresponding sorted index array would be $1, 0, 2$, because the tuples, after reordering the join column to index 0, follow the lexicographic ordering $(1,2,2) < (1,2,5) < (5,2,9)$. In essence, the sorted index array decouples the sorting order of tuples from their physical arrangement in the data array. It maintains the positions of tuples within the data array, and these positions are sorted based on the lexicographical order of their associated tuples.
% As a result, even if tuples with similar join column values, such as $(5, 2, 0)$ and $(5, 2, 9)$, are not stored contiguously in the data array, their corresponding positions are closely positioned in the sorted index array.

\paragraph*{Construction}
We extensively use NVIDIA's Thrust library \cite{thrust} to perform tasks such as copying, gathering, and sorting. The first step in the process of creating the sorted index array is to reorder the columns, which we perform using Thrust's transform function. This can be seen in line numbers $1$ to $5$ in the Algorithm~\ref{alg:index-construct}. After the reordering phase, we populate the sorted index array, by using Thrust's stable sort utility. This phase performs a stable sort based on the least significant column (rightmost) of the tuple and progresses towards sorting by the most significant column (leftmost). This sorting process is similar to radix sort; however, in this case, tuples are sorted one column at a time (rather than one bit at a time). This can be seen in line numbers $7$ through $10$.

\paragraph*{Fast range-queries}
Rearranging the column order to position the join columns at the beginning and then sorting the tuples offers two significant benefits. First, it enables efficient range queries, and second, it facilitates the fast merging of two relations. While fast-range queries are crucial for performing {join} operations, merging relations is an essential part of semi-na\"ive evaluation. As discussed in Section~\ref{sec:datalog}, the new relation is merged into the full relation at each iteration of the fixed point.

The sorted-index array works in tandem with the hash-map (described in the next subsection) to execute range queries. The hash map directly positions a GPU thread (in O(1) time) at the appropriate location within the sorted index array. Since the sorting process groups together tuples with identical join-column values, a range query can easily retrieve the entire set of tuples sharing the same join-column values by a linear scan.

\paragraph*{Deduplication}
\label{sec:impl-data-insert}
HISA  supports deduplication [R4] by first sorting all columns in lexical order, facilitating the rapid identification and removal of duplicates. The deduplication process is then executed by comparing each tuple with its adjacent tuple during a parallel scan. This approach ensures both efficient grouping for joins and effective deduplication, leveraging the parallel processing capabilities of the GPU.

\paragraph*{Merge operation}
\revised{Parallel merging of two sorted arrays is a common pattern in GPGPU programming. We utilize the path merge algorithm~\cite{green2012merge} provided by NVIDIA’s Thrust library to merge the sorted index arrays of two HISAs.}

\subsection{Open-Addressing Hash Table}
The sorted index array stores tuples in increasing order, clustering tuples which share the same join-columns.
To further optimize range queries, enabling retrieval of all tuples sharing the same join columns and meeting requirement [R2], HISA incorporates an open-addressing-based hash table. Using a hash table avoids control flow divergence and improves memory access efficiency compared to other range query techniques, such as tree-based searches.

The hash table is constructed to store distinct hash values (as keys) derived from the tuples of the data array, while only considering the join-columns to compute the hash. These keys are then associated with the smallest index of a tuple in the sorted index array (as values) that contains the corresponding join-column values.
An example of this hash table can be seen at the bottom of Figure~\ref{fig:hisa}. In this example, the hash table entry on the far right captures the hash value of join-columns $(5,2)$ and associates it with the starting position of all tuples having the join-column $(5,2)$, indicated by the $127^{th}$ position in the sorted index array.
Instead of directly storing the join column values---which may exceed $64$ bits in size---we opt to store their hash value as the key; this strategy effectively meets requirement [R3].

\begin{algorithm}[t]
\caption{Hash table construction algorithm}
\label{alg:hashtable-construct}
\begin{algorithmic}[1]
\For{each tuple $T$ in input relation, perform in parallel}
\State $H$ $\gets$ $hash(T)$
\State $I$ $\gets$ $H \pmod{\text{hash\_table\_size}}$
\State $T_{pos} \gets$ $\text{get\_position(H, sorted\_index\_array)}$
\While{$T$ is not inserted}
\State $K$, $V$ $\gets$ $\text{hash\_table[I]}$
\State $K' \gets$ AtomicCAS($K$, $H$)
\If{$K'$ $=$ $H$ or $(K,V)$ is uninitialized}
\While{{$T_{pos} < V$}}
\State $V \gets$  AtomicCAS($V$, $T_{pos}$)
\If {$V = T_{pos}$}    
\State \textbf{break} \Comment{Inserted $T$ in $hash\_table$}
\EndIf
\EndWhile

% \While{true}
% \If{$T_{pos} < V$}
% \State $V \gets$  AtomicCAS($V$, $T_{pos}$)
% \If {$V = T_{pos}$}    
% \State \textbf{break} \Comment{Inserted $T$ in $hash\_table$}
% \EndIf
% \Else 
% \State \textbf{break}
% \EndIf
% \EndWhile
\Else
\State Perform linear probing on $I$
% \State $I$ $\gets$ $I+1$$\pmod{\text{hash\_table\_size}}$ 
% \State
% \Comment{ Collision: linear probing}
\EndIf
\EndWhile
\EndFor
\end{algorithmic}
\end{algorithm}

\paragraph*{Construction} 
The pseudo-code for the construction of the hash table can be seen in Algorithm~\ref{alg:hashtable-construct}. We construct an open-addressing hash table, employing linear probing to handle collisions. The keys of the hash table are generated from the data array, while the values are derived from the sorted index array. 
%GPU threads access all elements of the data array via iterating through the sorted index array and referencing to the date array (line 1), computing the hash for each set of join-column values (lines 2 and 3). These hash values serve as the keys for the hash table, and the corresponding values are the positions of those keys in the sorted index array (line 4). 
GPU threads go through the sorted index array, which allows them to access all elements of the data array. They calculate hash values for every tuple, using only the join-column values and use these hash values as keys in the hash table. The hash table's values store the positions of the corresponding keys in the sorted index array, enabling quick lookup and access to the relevant data elements.
To handle collisions that occur when two keys map to the same slot in the hash table, we employ linear probing. However, since insertion is performed in parallel by thousands of threads, it is possible that during collision resolution, two threads with hash collisions attempt to write to the same hash slot.
To address this challenge, we utilize the \texttt{atomicCAS} (Compare-and-Swap) operator at line 7 of the algorithm, which ensures that only one thread can successfully update a slot at a time.
%As is with any hash table, a collision occurs when two keys are mapped to the same slot in the table, to overcome this challenge we using linear probing, however, since insertion in our case is happening in parallel by thousands of threads, we must avoid race conditions when a thread overrides data from other threads. To solve this issue we extensively rely on the usage of atomicCAS operator.
%In instances where two hash values map to the same entry, the collision is resolved using a linear probing as can be seen in the while-loop, fom lines $5$ to $22$.
%Actual insertion of key-value pairs are performed using \texttt{atomicCAS} (atomic compare and swap)~\cite{} operator to ensure that threads do not override each other.
%that ensuresin the has tables 
%which searches for the next available hash table slot.

% The way we have designed our hash table, among the tuples with the same join-column values, we only store the position of the smallest one.
In parallel execution, multiple GPU threads update smallest index of a tuple in the sorted index array simultaneously. This also leads to a race condition when two tuples sharing identical join column values are concurrently inserted into the hash table by two separate GPU threads.  We handle this from line 9 to line 14 of the algorithm, also using \texttt{atomicCAS} operator, which updates a value only if the new key-value pair is smaller than one existing in the hash table.

To conduct a range query, the process begins with calculating the hash value of the join columns. Based on this hash value, the method navigates to the smallest tuple possessing the specified join-column values and then proceeds with a linear scan until encountering a tuple where the join-column value no longer matches. We show this in more detail in the context of a join operation in Section V.A.
% This mechanism enhances data fetching efficiency by leveraging the hash table for quick access to grouped tuples, significantly improving performance on GPU architectures by reducing the need for incoherent memory access and branching. 

\paragraph*{Merge operation}
\revised{Currently, hash table merging of HISA is done by inserting all elements from the Delta hash table into the Full table. For each existing key, the value is updated if the Delta value is smaller than the current value in the Full hash table. While we consider  more efficient algorithms (e.g., Cuckoo Hashing\cite{li2021dycuckoo}) as an important future direction, GPU hash tables can process up to 0.5 billion keys per second, making this approach faster than merging sorted arrays, which requires significant time for buffer allocation and deallocation during path-merging. }

\section{\tool{}: GPU-powered datalog engine}
%Using HISA for Datalog on the GPU}
\label{sec:gdlog}

In this section, we introduce \tool{}, our GPU-accelerated implementation of Datalog that leverages the HISA to optimize performance and efficiently utilize the GPU. We focus our discussion on three distinguishing aspects: (a) implementation of the semi-na\"ive evaluation in the fixed-point loop, (b) a novel optimization to significantly accelerate $n$-way joins: temporary materialization, and (c) eager buffer management, which leverages amortization to improve memory allocation.

\begin{figure}[]
    \centering
    \includegraphics[width=0.78\linewidth]{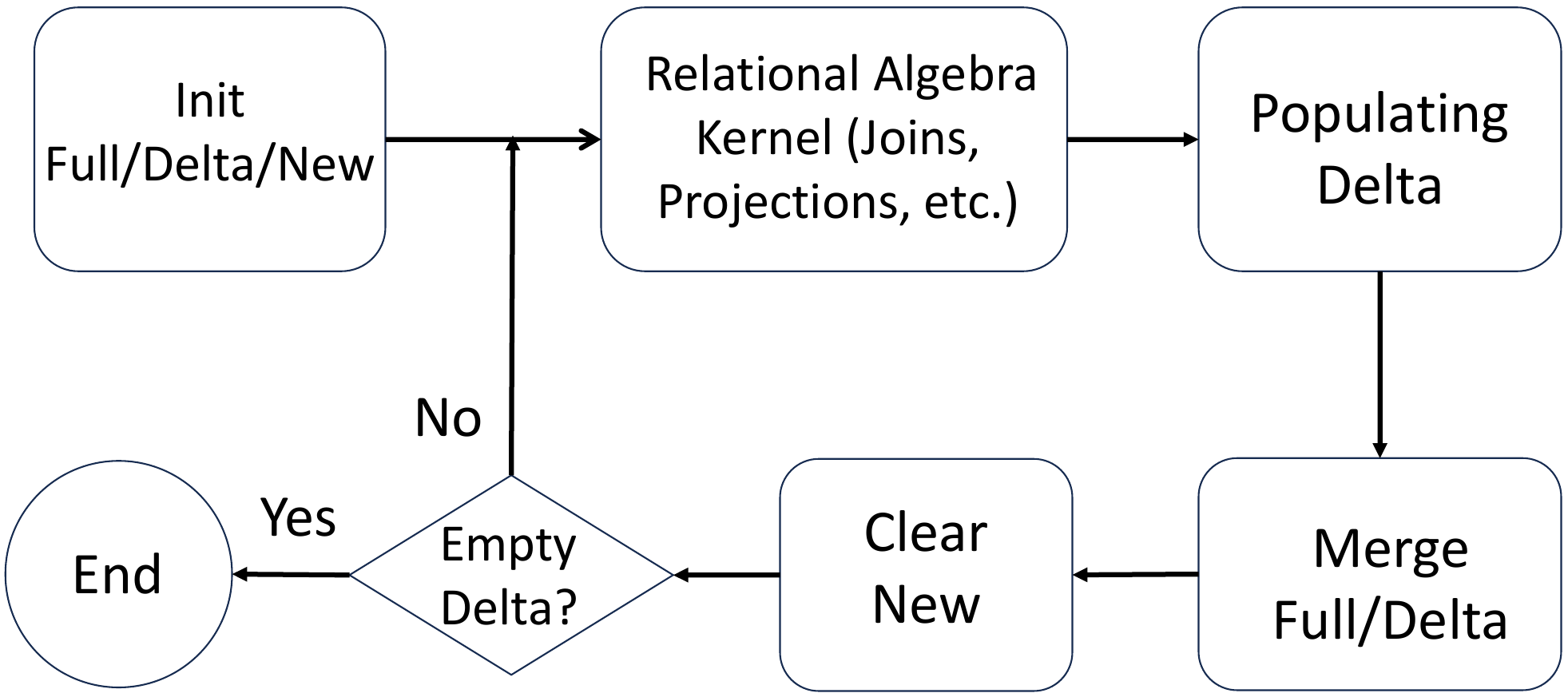}
    \caption{\fontsize{9}{11}\selectfont Workflow of semi-na\"ive evaluation.}
    \label{fig:eval-flow}
\end{figure}

\subsection{Evaluation of Datalog Programs}
\label{sec:gdlog-eval}
%GDLog implements the semi-naive evaluation workflow as described in Section~\ref{}. Figure~\ref{fig:eval-flow} delineates \tool's evaluation workflow, which comprises five distinct phases: "Updating Index", "Join", "Populate Delta", "Merge Full/Delta", and "Clear New".

Datalog's evaluation is often given in terms of relational algebra, which includes relational operators such as join and projection. Evaluation executes these kernels in a loop. In each iteration of the loop, compute kernels (compiled from the program rules) generate new tuples (facts); these newly-discovered facts may then trigger further deduction in subsequent iterations. The process continues until a fixed point is reached, where no new facts can be derived. Figure~\ref{fig:eval-flow} illustrates the detailed steps involved in this evaluation process, namely ``Join,'' ``Populate Delta,'' ``Merge Full/Delta,'' and ``Clear New.''

%\paragraph*{Updating Index}
%Within a Datalog program, it's common for each relation to be associated with multiple indexes due to the variety of ways it might be queried or joined. Taking the Same Generation query as an example, as described in Section~\ref{sec:datalog}, the \textit{SG} relation undergoes joins on both its first and second columns, thus requiring the generation of two distinct indexes for \textit{SG}. In essence, we create two HISA data structures for the edge relation, one indexed on the first column and the other indexed on the second column. 
%\tool{} ensures that all incremental changes to a relation are accurately reflected across all its indexes. 

\begin{figure}[t]
    \centering
    \includegraphics[width=1\linewidth]{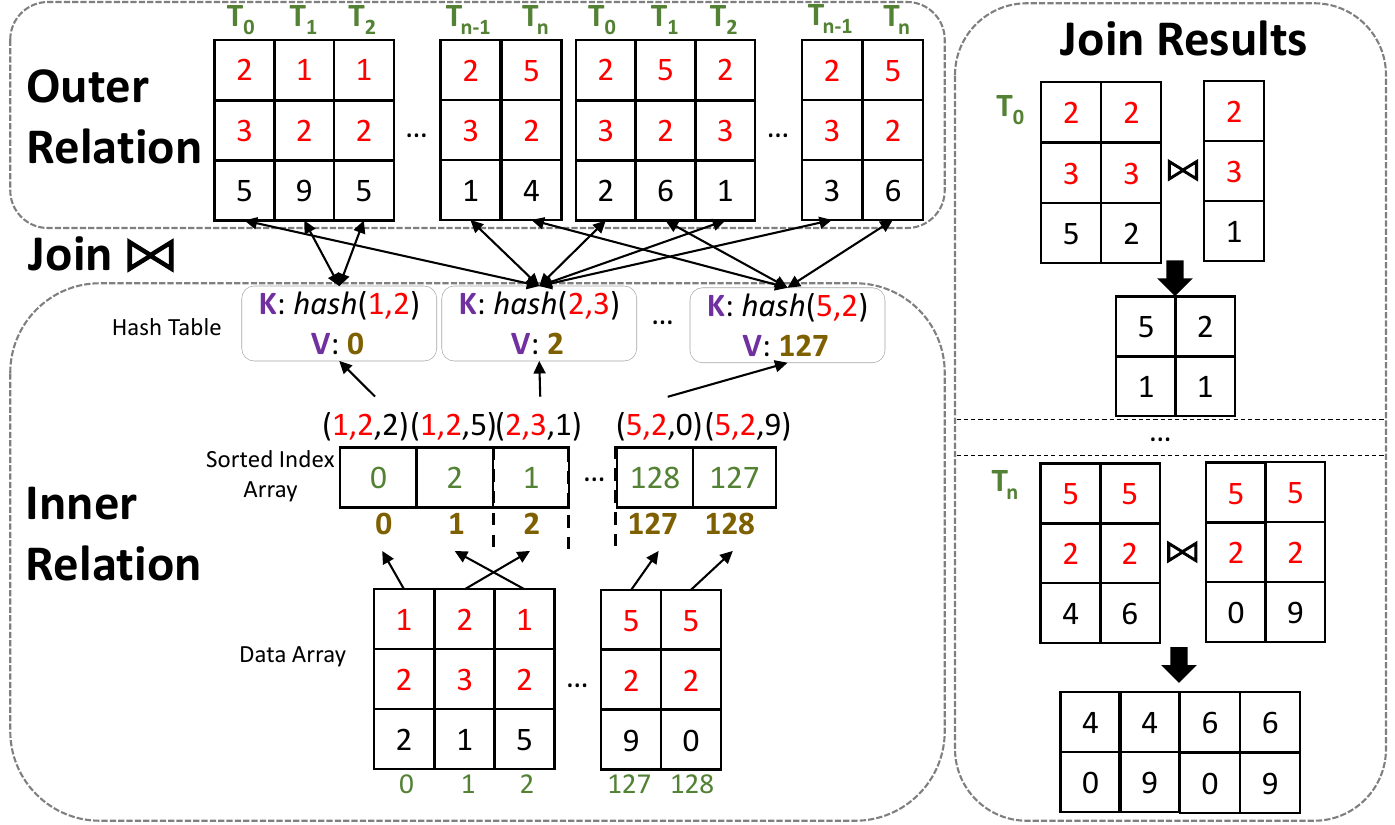}
    \caption{Example of a join between two 3-arity relations \textit{Bar} (inner) and \textit{Foo} (outer).}
    \label{fig:join_layout}
    % \vspace{-0.2cm}
\end{figure}

\paragraph*{Implementing Joins}
\label{sec:join-impl}
% The second step in the fixpoint computation involves the execution of join operations in \tool{}.
% Leveraging our hybrid data structure, these join processes are designed to harness the benefits of both hash and sorted joins, making them well-suited for recursive query scenarios on GPUs.
The bulk of the evaluation overhead of modern Datalog evaluation lies in iterated joins. For example, consider the following query:
%We illustrate the implementation of the \emph{join} operation, through following query: 
\[
\begin{array}{lcl}
    \textit{Foobar}(c,d)  & \leftarrow &  \textit{Foo}(a, b, c), \textit{Bar}(a, b, d). 
\end{array}
\]
This join operation involves two relations, each with three attributes, sharing the first two columns as join columns in Figure~\ref{fig:join_layout}.
To facilitate parallel iteration over all tuples of the outer relation, we use the data array component of the HISA data structure.
As shown in the figure, we access the data array of the outer relation in \emph{stride} units. The size of each stride corresponds to the total number of threads employed for concurrent execution. 
Each GPU thread concurrently retrieves a tuple within a stride, where the thread ID corresponds to the tuple's offset within that stride.
In the example shown all tuples of the outer relation are accessed in two strides, the first stride ranges from tuples $\textit{Foo}(2,3,5)$ to $\textit{Foo}(5,2,4)$ and the second stride ranges from $\textit{Foo}(2,3,2)$ to $\textit{Foo}(5,2,6)$. We encapsulate this strided access in the form of a parallel for in line number $1$ of the Algorithm~\ref{alg:join}.
Users can configure the stride size based on their GPU, with a recommended size being $32$ times the number of stream processors.
By accessing the data in the outer relation using this stride-based approach, data locality is improved, leading to optimal utilization of the cache and enhancing overall performance.

The lower section of the figure demonstrates how each thread queries the inner relation.
The range queries are facilitated by using the hash table and the sorted index array components of HISA. 
The hash table is used to find the starting position of the relevant queries in the sorted index array (lines $2$ to $5$), which is then linearly scanned to retrieve all matching tuples from the data array (lines $6$ to $10$).
For instance, thread $T_0$ operates on the outer relation tuple $Foo(2,3,5)$ and $Foo(2,3,2)$.
After hashing column $(2,3)$, \tool{} uses the hash table of the HISA structure to find the smallest tuple at position $2$ in the \textit{Bar} relation's sorted index array, yielding $\textit{Bar}(2,3,1)$. A scan starting from this position reveals that the only matching tuple in the \textit{Bar} relation is $\textit{Bar}(2,3,1)$. On the right side of the figure, the join results show that tuples $\textit{Foobar(5,1)}$ and $\textit{Foobar(2,1)}$ are generated as a result of the join between matched tuples from the outer and inner relations.

\revised{
Although both employ hash join-like algorithms, \tool{} differs from the join algorithm used in another Datalog engine prototype, GPUJoin~\cite{shovon2023towards}. In GPUJoin, the hash table directly contains all tuples, and during the join, tuples are accessed through linear probing of the hash table. This design can lead to larger hash tables and increased memory overhead, especially when a low load factor is used for faster construction. In contrast, our tool’s join algorithm accesses the hash table only to find the starting position of tuples sharing the same indexed column. This approach keeps the hash table size small and allows for fast construction.
}

\begin{algorithm}[t]
\caption{Parallel Binary Join on GPU}
\label{alg:join}
\begin{algorithmic}[1]
\For{$T_{outer}$ in $\textit{Relation}_{outer}$ \textbf{parallel}}
    \State $\textit{ht} ~ \gets$ hash($T_{outer}$.join\_columns)
    \State $\textit{index\_pos} \gets \textit{ht} ~ \% ~$ hash table size of $ \textit{Relation}_{inner}$
    \State $\textit{pos} \gets \textit{Relation}_{inner}$.sorted\_index\_array[\textit{index\_pos}]
    \State $T_{inner} \gets$ tuple at \textit{pos} in $\textit{Relation}_{inner}$.data\_array
    \While{$T_{inner}.$join\_columns $\neq T_{outer}$.join\_columns}
        \State generate result tuples based on $T_{outer}$ and $T_{inner}$
        \State \textit{pos}++
        \State $T_{inner} \gets$ tuple at \textit{pos}
    \EndWhile
\EndFor
\end{algorithmic}
\end{algorithm}

\paragraph*{Populating delta}
\label{sect:pop-delta}
The next step in the fixpoint loop involves populating the \textit{delta} relation, which will be used in the join phase of the next iteration. \revised{In this step, \textit{delta} is constructed by removing from \textit{new} the tuples that are already present in \textit{full}.} \tool{} accomplishes this using a relational algebra ``set difference'' applied to the \textit{new} and \textit{full} relations.
It is worth noting that some prior work, such as GPUJoin, fuses this step with the merging step by directly merging the non-deduplicated \textit{delta} relation and then deduplicating the merged \textit{full} relation. This fusion approach is efficient when the size of the \textit{full} relation is small. However, deduplicating the \textit{full} relation requires a full scan of all the tuples in the \textit{full} relation, which becomes highly expensive as the size of the \textit{full} relation grows. Therefore, \tool{} takes a different approach by separating the \textit{delta} relation population into a distinct phase.

\paragraph*{Merging Full with Delta and Clearing New}
The final step before fixpoint checking involves merging all the tuples within the \textit{delta} relation into the \textit{full} relation, followed by the removal of all tuples within the \textit{new} relation. Leveraging the HISA bulk insertion technique outlined in Section~\ref{sec:impl-data-insert}, we incorporate all deduplicated new tuples from \textit{delta} into \textit{full} in parallel on the GPU.
It is worth noting that the management of buffer memory during merging between \textit{full} and \textit{delta} plays a crucial role in ensuring the optimal performance of larger queries. %Subsequently, we will delve into the finer details of this memory management strategy in the following subsection.

% As we have previously discussed in the background and motivation ~\shovon{should we use ref?} section, this particular phase is inherently serial and has the potential to become a bottleneck when dealing with larger query sizes in CPU-based Datalog engines. In \tool{} we can take advantage of GPU operation over our data structure's internal compact array to parallelize this process.
% Efficient parallelization is achieved by employing a strategy that entails the coarse yet swift partitioning of both the sorted array of \textit{full} and \textit{delta} relations into several small, sorted tiles, each of which can fit neatly into a GPU's execution wraps. This transformation of the merging process effectively converts it into a divide-and-conquer problem and, as a result, makes it amenable to seamless acceleration by SIMD hardware.

\subsection{Temporarily-Materialized $n$-way Joins}
\label{sec:impl-k-way}
% The second type of trade-off occurs when more than 2 relations are involved in a join. 
While simple Datalog queries use only binary joins, practical applications include $n$-way joins. An example is the Same Generation (\emph{SG}) query  mentioned in Section~\ref{sec:datalog}. 
The second rule of \textit{SG} represents an inductive case that necessitates joining \emph{SG} with \emph{Edge} twice. When employing semi-na\"ive evaluation, its join plan may be expressed via the following relational algebra operation:
\begin{displaymath}
\begin{array}{lcl}
        \textit{SG}_{\emph{new}} & = & \textit{Edge}_{\emph{full}} \bowtie \textit{SG}_{\emph{delta}} \bowtie \textit{Edge}_{\emph{full}} 
    \end{array}
\end{displaymath}
Various mechanisms exist for evaluating $n$-way joins; while special-purpose joins (such Leapfrog Triejoin~\cite{veldhuizen2014leapfrog}) can achieve worst-case optimal bounds, they do so by imposing representation restrictions. Modern engines tend to favor nested-loop joins~\cite{jordan2016souffle}, in which case $n$-way joins are ordered into sequences (or trees) of binary joins, with an emphasis on optimizing the join order\cite{zhao2024evaluating}. In the case of \emph{SG},  it first loops over  $\texttt{Edge}_{\emph{full}}$ and within the loop body, it further loops on all matched  tuples in $\texttt{Edge}_{\emph{delta}}$. This generates another nested loop over to join with $\texttt{Edge}_{\emph{full}}$.
% there are three distinct methods (unique combinations of 3 join candidates) for dividing the join into sub-joins and allocating them across threads.  For example, one strategy involves partitioning the entire join based on tuples in the outermost relation (\emph{i}.\emph{e}., tuples in $\texttt{SG}_{\emph{delta}}$).

\begin{figure}[t]
    \centering
    \includegraphics[width=0.9\linewidth]{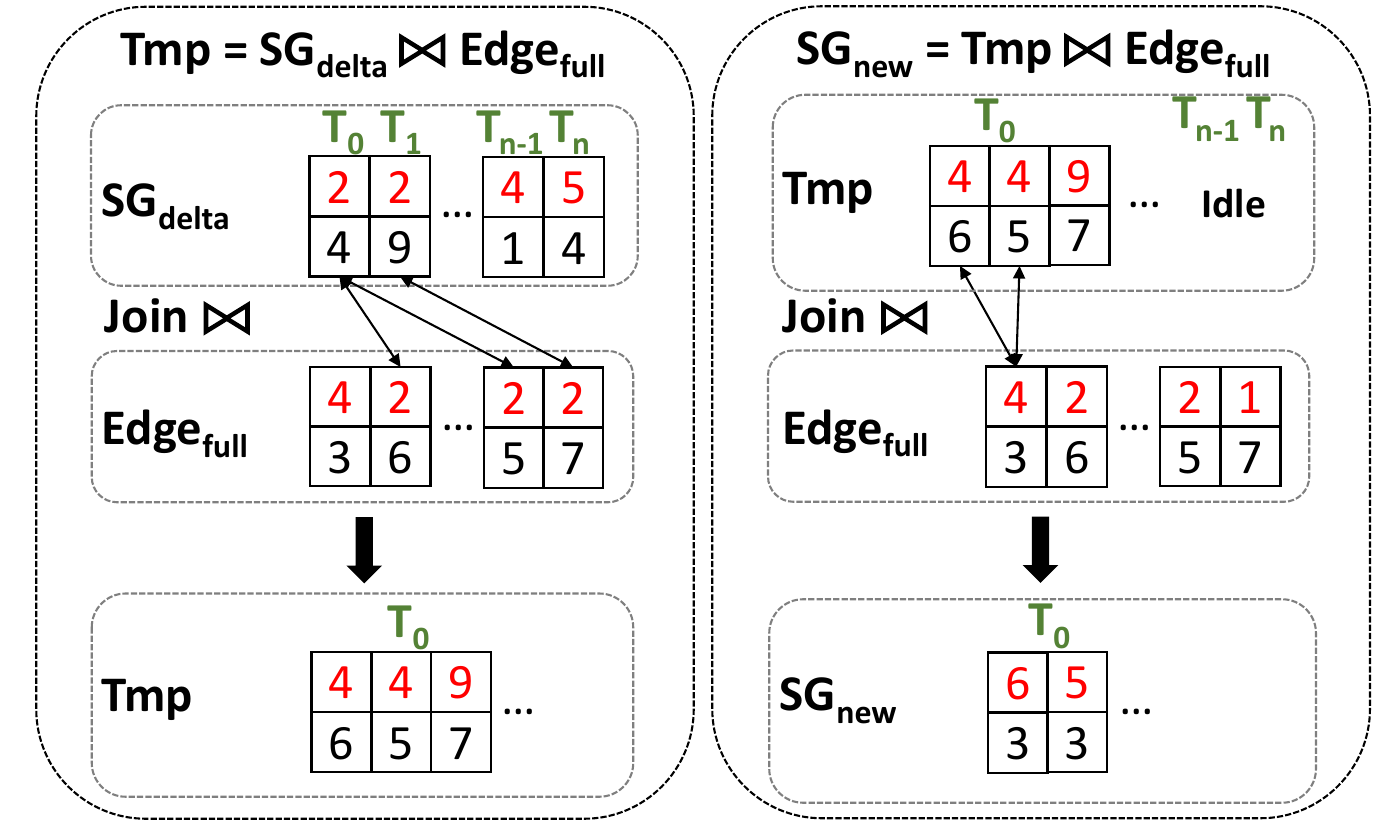}
    \caption{Thread waiting in nested non-materialized join due to workload imbalance. $T_0$ to $T_n$ are in the same warp.}
    \label{fig:join_imbalance}
\end{figure}

% Multi-core CPUs operate on a Many Instruction Many Data (MIMD) architecture, allowing each thread to execute different instructions concurrently without experiencing idle time caused by conditional branching in the program. However, this workload division approach doesn't align with the hardware nature of GPUs. GPUs utilize a Single Instruction Many Data (SIMD) architecture, where threads within the same GPU warp can only execute the same instruction at any given time.
Due to the architecture of GPU, conditional branching in the program can lead to threads idling during an $n$-way join.
For example, in Figure~\ref{fig:join_imbalance}, thread $T_0$ is assigned to compute the join result for the tuple $\texttt{SG}(2,4)$. During the first join with $\texttt{Edge}_{\emph{full}}$, it generates numerous temporary results. Unfortunately, thread $T_{n-1}$ and thread $T_{n}$ lie within the same GPU warp and do not find any matching tuples. In the subsequent join between the first set of join results and $\textit{Edge}_{\emph{full}}$, $T_0$ must continue the computation, while threads $T_{n-1}$ and $T_n$, have to remain idle until $T_0$ completes the second join and returns.

Another approach to divide the workload among threads involves materializing each sub-join. We leverage this observation by using a ``temporary materialized'' buffer. In \textit{SG}, the second rule can be seen as two sequential binary joins:
\[
    \begin{array}{lcl}
        \textit{Tmp}(b, x) & \leftarrow & \textit{Edge}(a,x), ~ \textit{SG}(a, b).  \\
        \textit{SG}(x, y) & \leftarrow & \textit{Edge}(b, y), ~ \textit{Tmp}(b, x).
    \end{array}
\]
In relational algebra, an explicit join plan may be written as:
\[
    \begin{array}{lcl}
        \textit{Tmp} & = & \textit{SG}_{\emph{delta}} \bowtie \textit{Edge}_{\emph{full}} \\
        \textit{SG}_{\emph{new}} & = & \textit{Tmp} \bowtie \textit{Edge}_{\emph{full}}
    \end{array}
\]
This split does not make a significant impact on performance for CPU based system. However, in GPU-based systems, having two sequential joins scheduled into two separate executions of GPU kernel functions is notably faster from the first method, where only one GPU kernel function executes. By utilizing the technique of temporal materialization, data parallelism is optimized to its fullest extent. This optimization ensures that all available threads are actively engaged in computation, effectively eliminating any idle time.

The first join's workload is based on tuples in $\textit{SG}_{\emph{delta}}$. By contrast, the workload of the second join is \revised{divided} based on the $\textit{Tmp}$  tuples. In the previous example, temporary join result tuples like $\textit{Tmp}(9, 7)$ are assigned to other idle threads, such as thread $T_{n-1}$, $T_n$, rather than all being processed by the busy thread $T_0$. In this way, all threads in the same GPU wrap will have a similar workload, eliminating idle threads.

%By utilizing the technique of temporal materialization, data parallelism is optimized to its fullest extent. This optimization ensures that all available threads are actively engaged in computation, effectively eliminating any idle time.

% The main drawback to temporarily-materialized joins is the extra space overhead required  to store the temporary relations. However, unlike in shared-nothing approaches (such as BPRA~\cite{kumar2020load})---wherein intermediate results must be persistently materialized in space---our approach purges temporarily tuples once the associated join completes.
% Additionally, unlike normal relations, temporarily-materialized results do not require multiple versions, sorting, or deduplication.

\subsection{Buffer Management and Amortized Allocation}
\label{sec:impl-buffer}

\revised{As we will see in Section~\ref{sec:eval-buffer}, the merge operation of \emph{delta} tuples into the \emph{full} version of the relation becomes a bottleneck, consuming up to 45\% of the total execution time. Buffer management during the merge process is particularly expensive, as it requires reallocating new buffers and copying data equivalent to the combined size of the \emph{full} relation and the \emph{delta} for each iteration. To address this, \tool{} employs a strategy called Eager Buffer Management (EBM).
Unlike the buffer management policies in other GPU Datalog engines, such as GPUJoin, the buffer allocated using EBM in \tool{} is not freed immediately after the merge operation. Before performing the merge, \tool{} checks if the buffer allocated in the previous iteration is sufficient for the current merge operation. If it is, the buffer is reused; if not, rather than allocating a buffer with the exact size of the full and delta tuples, \tool{} allocates a buffer sized to the full tuple size plus \textit{k} times the delta tuple size, where  $k$  is a tunable parameter depending on the total size of VRAM. This strategy reduces the overhead of frequent buffer allocations in every iteration and uses a small amount of extra memory to enable faster iterative computation.}

\section{Evaluation}
\label{sec:eval}

In this section, we conduct a comprehensive performance evaluation of \tool{}, demonstrating the impact of our optimization techniques. Our assessment begins by analyzing how eager buffer management influences both query execution time and system memory usage.
% Subsequently, we dig into a comparative study highlighting the substantial advantages of our balanced k-way join approach over the nested k-way join methodology when executed on a GPU.
We then compare the performance of \tool{}  against existing Datalog engines, utilizing well-established Datalog queries and real-world datasets. Due to the limited availability of mature GPU-based Datalog engines and the inaccessibility of GPUDatalog \cite{martinez2013Datalog} in both source and binary forms, we also include a state-of-the-art CPU-based system (Souffl\'e) in our comparison.  Although DCDatalog \cite{wu2022dcdatalog} recently claims to have surpassed Souffl\'e in speed, the paper's artifacts are not accessible for evaluation. Finally, we demonstrate the practicality of \tool{} for program analysis, showing significant performance boosts over CPU-based solutions.

\subsection{Environment Setup}
% Before diving into the details of our evaluation, we provide a brief overview of our experimental environment settings:

\paragraph*{\tool{} and Other GPU-Based Systems}
% All our CUDA-related code runs on a server equipped with a 32-core Intel Xeon Gold 6338 CPU, operating at 2.00GHz, backed by 128GB of DRAM, and features one NVIDIA A100 80GB PCIe GPU.
Except for Section~\ref{sec:protablility}, all our CUDA-related code runs on a server equipped with a 64 core AMD EPYC 7713 CPU (code name Milan), and an NVIDIA H100 80GB GPU. \yihao{The CUDA toolkit version we use is 11.8. For cuDF, we employ version 23.10, the latest version available at the time of our evaluation, also running under CUDA 11.8.} The code for Datalog queries in cuDF and GPUJoin is sourced from its public GitHub repository~\cite{gpujoincode}. In Section~\ref{sec:protablility}, our \tool{}-HIP implementation is tested on 2 machines. One has dual AMD EPYC 7713 processors together with four 64GB AMD Instinct MI250 cards. Another has dual AMD EPYC 7742 and four 32GB AMD MI50 GPUs. In our benchmark, only one GPU for each system is used. The NVIDIA A100 results are tested on a machine with a 32-core Intel Xeon Gold 6338 CPU. \yihao{Although the testbed we rented utilizes high-end CPUs, it is important to note that using lower-performance CPUs does not affect our GPU tool's performance. We validated our A100 results on a less powerful CPU and obtained nearly identical results.}

\paragraph*{CPU Based System: Souffl\'e}
Our Souffl\'e experiments are conducted on a single node (1$\times$AMD EPYC 7543P) with 512G RAM. All Souffl\'e queries are compiled into C++ code \revised{and then compiled using gcc-10 with the optimization flag -O3. To fully utilize all 32 cores on the test platform’s CPU during compilation, we used the -j 32 option with the build system} . We found this setting delivered the best runtime on our test platform.

% We have Souffl\'e emit its join plan via the flag \texttt{--show=transformed-ram}. We then ensure that all tools used in our experiments employ the same join plan as Souffl\'e by manually replicating Souffl\'e's plan in our implementations using \tool{}, cuDF, and GPUJoin.

\subsection{Test Programs and Datasets}

We evaluate reachability (\textit{REACH}, Section I), same generation (\textit{SG}, Section II), and context-sensitive program analysis (\textit{CSPA})~\cite{lam2005context}. We include \textit{REACH} as a common baseline which stresses iterated binary joins, without any need for temporal materialization. We include \textit{SG} and \textit{CSPA} as queries which require more intelligence in join planning and stresses support for $n$-way joins. Recall the code for \textit{SG} was introduced in Section~\ref{sec:impl-k-way}. % We include \textit{CSPA}, a 10-rule context-sensitive points-to analysis, to evaluate \tool{}'s potential for impact in high-throughput program analysis. We use data from RecStep~\cite{recstep}, and its Datalog code is presented in Figure~\ref{fig:cspa-code}.
We ran our experiments using ten large graphs for both \textit{REACH} and \textit{SG}. These graphs were chosen from the Stanford Network Analysis Project network dataset, SuiteSparse, and road network datasets~\cite{snapnets, suitesparse, li2005trip}, and vary in size and application domain. For \textit{CSPA}, we used input data provided by the authors of RecStep~\cite{recstep}.
% and used in evaluation of a program-analysis tool named Grapan~\cite{Wang2017GraspanAS}.
% These facts were extracted using method cloning to include calling-context information from three distinct open-source applications: the Linux kernel, httpd, and PostgreSQL. 

% \vspace{-0.1cm}
\subsection{Evaluating \buffer}
\label{sec:eval-buffer}
We assess the efficacy of our \buffer (EBM) mechanism by conducting a comparative analysis of runtime and memory usage with EBM enabled and disabled, using the \textit{REACH} query on five distinct large graphs. 
Our results are presented in Table~\ref{tab:buffer_compare}. This table is organized into four groups of columns. The first group specifies the datasets used in the experiment. The second group of columns provides insights into the iteration counts within the \textit{REACH} query. This group is further divided into two sub-columns. ``Total'' denotes the total
number of iterations, while ``Tail'' refers to the count of tail iterations, where the number of \textit{delta} tuples generated in that iteration is less than 1\% of the total tuples in the resulting \emph{Reach} relation. Notably, the absence of a tail iteration number for the ``usroads'' dataset is due to the fact that the number of \textit{delta} tuples generated in every iteration is less than $1$\%. The third group of columns displays the running time and speed-up ratio achieved by enabling Eager Buffer Management (EBM). It reveals that EBM accelerates all \textit{REACH} queries, with greater benefits in longer iterations, especially long tail ones. ``\textsf{usroads}'' benefits the most, achieving a 3$\times$ acceleration. This is because relations that have a higher percentage of tail iterations are bound to benefit more from EBM, as the extra buffer allocated as part of the EBM will be able to accommodate the buffer merge requirements of a large fraction of the tail iterations, requiring fewer memory reallocations.

\begin{table}[t]
    \fontsize{9}{11}\selectfont
    \caption{\fontsize{9}{11}\selectfont Comparing runtime and memory usage of \textit{REACH} in \tool{} with and without eager buffer management on NVIDIA H100}
    \begin{tabular}{@{}c||cc||cc||cc@{}}
   \toprule
    & \multicolumn{2}{c|}{Iterations}  & \multicolumn{2}{c|}{Query Time (s)}  & \multicolumn{2}{c}{Memory(GB)}   \\
    \midrule
             Dataset   &  Total & Tail  & Normal        & Eager       & Normal         & Eager                  \\
    \midrule 
    usroads     & 606  & /  & 52.42         & 17.53       & 20.35          & 26.84         \\
    vsp\_finan  & 520  & 491  & 59.08         & 21.91       &  20.22          & 28.26       \\
    fe\_ocean   & 247  & 90   & 47.19         & 23.36       &  37.97          & 50.43        \\
    com-dblp    & 31   & 18   & 17.83         & 14.30       &  43.24          & 60.18       \\
    Gnutella31  & 31   & 17   & 4.80          & 3.76        & 20.22          & 28.26        \\
    % fe\_sphere  & 2.19          & 1.69        & 1.29x    & 2.29           & 2.88          & 1.25x    \\ 
    \bottomrule 
    \end{tabular}
   \label{tab:buffer_compare}
\end{table}

\begin{table}[h]
    \fontsize{9}{11}\selectfont
    \caption{\fontsize{9}{11}\selectfont Reachability execution time comparison: \tool{} (NVIDIA H100) vs. Souffl\'e (AMD Milan CPU 32 cores), GPUJoin, and cuDF on large graphs (OOM: out of memory).}  
    \begin{tabular}{c||c||c|c|c|c}
    \toprule
    \multicolumn{1}{c}{Dataset} & \multicolumn{1}{c}{\textit{Reach}} & \multicolumn{4}{c}{Time (s)} \\ 
    \multicolumn{1}{c}{name} & \multicolumn{1}{c}{edges} & \tool{}      & Souffl\'e     & GPUJoin  & cuDF  \\ \midrule
    com-dblp            & 1.91B    & \textbf{14.30}     & 232.99   &  OOM    & OOM \\
    fe\_ocean           & 1.67B    & \textbf{23.36}     & 292.15   &  100.30 & OOM \\
    vsp\_finan          & 910M      & \textbf{21.91}     & 239.33   &  125.94 & OOM \\
    Gnutella31          & 884M      & \textbf{5.58}      & 96.82    &  OOM    & OOM \\ 
    fe\_body            & 156M      & \textbf{3.76}      & 23.40    &  22.35  & OOM \\
    SF.cedge            & 80M       & \textbf{1.63}      & 33.27    &  3.76  & 64.29 \\ 
    \bottomrule
    \end{tabular}
   \label{tab:reach_compare}
\end{table}

\subsection{\tool{} vs. SOTA GPU joins}

\label{eval:vs-other-joins}

In this section, we evaluate the performance of the \textit{REACH} query using \tool{}, GPUJoin, and cuDF, with Souffl\'e serving as the baseline. The results are summarized in Table~\ref{tab:reach_compare}.
The datasets used in the experiments represent a diverse array of sources, ensuring a comprehensive assessment that spans scientific communities, P2P networks, random graphs, and road networks. This dataset diversity helps mitigate any bias towards specific graph categories.

\begin{table}[t]
    \fontsize{9}{11}\selectfont
    \caption{\fontsize{9}{11}\selectfont Same Generation (\emph{SG}) execution time comparison: \tool{} vs. Souffl\'e  and cuDF. Souffl\'e running \revised{32 core AMD Milan CPU}; \tool{} and cuDF running on NVIDIA H100 GPU.}   
    % \begin{tabular}{lllll}
    \begin{tabular}{c||c||c|c|c|c}
    \toprule
    \multicolumn{1}{c}{\multirow{2}{*}{Dataset}} & \multicolumn{1}{c}{\multirow{2}{*}{SG size}} & \multicolumn{3}{c}{Time (s)} \\ 
    \multicolumn{1}{c}{}                         & \multicolumn{1}{c}{}                         & \tool{} & HIP      & Souffl\'e     & cuDF    \\ \midrule
    fe\_body                                     & 408M                                  & \textbf{5.05} & 19.57     & 74.26       & OOM    \\
    loc-Brightkite                               & 92.3M                                  & \textbf{3.42} & 14.00      & 48.18       & OOM        \\
    fe\_sphere                                   & 205M                                  & \textbf{2.36} & 8.48      & 48.12       & OOM        \\
    SF.cedges                                     & 382M                                   & \textbf{5.54} & 20.57      & 68.88       & OOM        \\
    CA-HepTH                                     & 74M                                   & \textbf{2.79} & 5.92      & 20.12       & 21.24        \\
    % delaunay\_n16 & 26M & \textbf{0.89} & 6.68  & 14.83 \\
    ego-Facebook  & 15M & \textbf{1.23} & 2.81 & 17.01 & 19.07 \\
    \bottomrule
    \end{tabular}
   \label{tab:sg_compare}
\end{table}

\tool{} consistently emerges as the fastest engine in all test cases. Compared to GPUJoin, \tool{} exhibits a remarkable speedup, with ratios often exceeding 3x. In the  ``fe\_body'' testcase, the speedup ratio reaches an impressive 6x. Notably, GPUJoin, designed specifically for reachability queries, stores output relations in raw arrays. This specialization suggests that \tool{} could achieve even higher speedup ratios when the method of GPUJoin is applied to general  queries.
Compared to GPUJoin, \tool{} used less memory: GPUJoin OOM'd twice during testing. This is because of its of use open-addressing-based hashmaps for storing relations, necessitating a low hashtable loading factor to facilitate joins.  By contrast, HISA employs hash tables only for accelerating range fetching on indexed relations. This results in efficient access even with a higher load factor ($0.8$).

%In contrast with \tool{}, cuDF's implementation yields unsatisfactory results, even when compared to the baseline CPU-based approach. It's complex code base makes it challenging to perform a detailed memory profile, but the presence of numerous out-of-memory errors during experiments points to memory-related performance issues. High memory usage significantly affects memory allocation and deallocation for buffer and tuple storage, contributing to slow running times.

\begin{figure*}[t]
\begin{tabular}{ccc}
\begin{minipage}{.46\textwidth}
    \captionof{table}{\fontsize{9}{11}\selectfont Context-Sensitive Program Analysis \emph{(CSPA)} execution time comparison: \tool{} (NVIDIA H100) vs. Souffl\'e (\revised{32 core AMD Milan CPU}); input data from~\cite{awad2019engineering}.}
    \label{tab:cspa_compare} 
\resizebox{\textwidth}{!}{\begin{tabular}{@{}c||c|c||cc||c@{}}
    \toprule
    Dataset    & Input Relation  & Output Relation  & \multicolumn{2}{c}{Time (s)}    & Speedup \\
    Name       & Size            & Size             & \tool{} & Souffl\'e              &      \\ \midrule
    Httpd      & \begin{tabular}[c]{@{}c@{}}Assign: 3.62e5\\ Dereference: 1.14e6\end{tabular} & \begin{tabular}[c]{@{}c@{}}ValueFlow: 1.36e6\\ ValueAlias: 2.34e8\\ MemAlias:8.89e7\end{tabular} &\textbf{1.33}         & 49.48   & \textbf{37.2x}   \\ \midrule
    Linux      & \begin{tabular}[c]{@{}c@{}}Assign: 1.98e6\\ Dereference: 7.50e6\end{tabular} & \begin{tabular}[c]{@{}c@{}}ValueFlow: 5.50e6\\ ValueAlias: 2.23e7\\ MemAlias:8.84e7\end{tabular} & \textbf{0.39}        & 13.44   & \textbf{34.5x}    \\ \midrule
    PostgreSQL & \begin{tabular}[c]{@{}c@{}}Assign: 1.20e6\\ Dereference: 3.46e6\end{tabular} & \begin{tabular}[c]{@{}c@{}}ValueFlow: 3.71e6\\ ValueAlias: 2.23e8\\ MemAlias:8.84e7\end{tabular} & \textbf{1.27}       & 57.82   & \textbf{44.9x}   \\
    \bottomrule
    \end{tabular}}
\end{minipage}
& 
\begin{minipage}{0.02\textwidth} % This is the spacer minipage
\hfill % Optional: use \hfill to adjust spacing further
\end{minipage}
&  
\begin{minipage}{.46\textwidth}
    \centering
    \includegraphics[width=\linewidth]{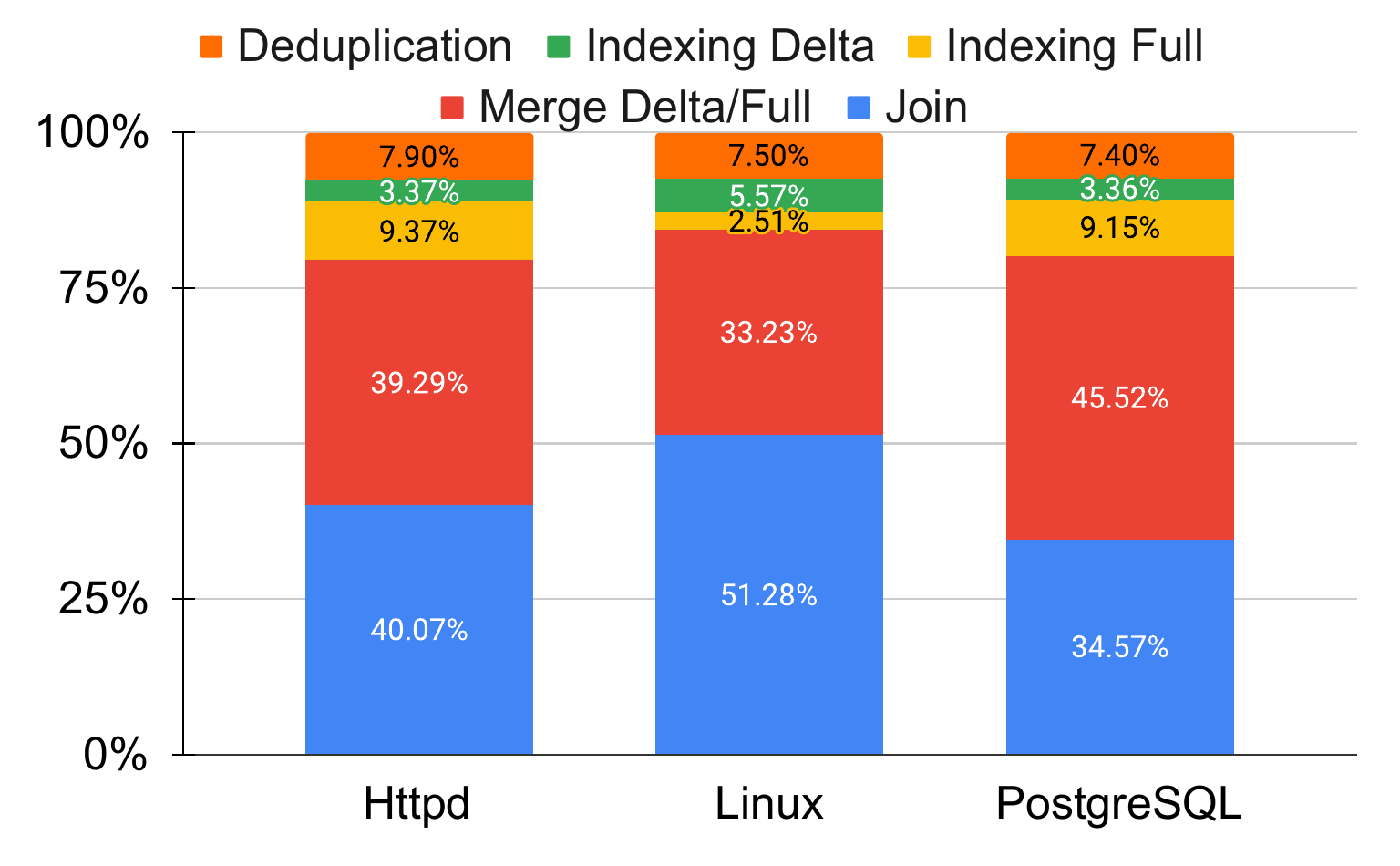}
    \caption{\fontsize{9}{11}\selectfont Running time detail analysis of \textit{CSPA} query using \tool{} on various real world datasets on NVIDIA A100.}
    % \caption{\revised{\fontsize{9}{11}\selectfont Running time detail of \textit{REACH} query using \tool{} and GPUJoin on fe\_body datasets on NVIDIA H100.}}
    \label{fig:join_detail}
\end{minipage}
\end{tabular}
\end{figure*}

We present another test case, \textit{SG}, to further validate our previous discussions. The results are presented in Table~\ref{tab:sg_compare}. The \textit{SG} query involves $n$-way joins and represents more general use cases. As GPUJoin does not support the SG query, we exclusively compare \tool{} with cuDF. Across all six test cases, cuDF encounters out-of-memory (OOM) errors four times, while the non-OOM test cases, although approaching the baseline performance of Souffl\'e, remain considerably slower than \tool{}. \tool{} consistently delivers stable performance, running nearly 7$\times$ faster than the baseline in most of cases, while not experiencing any OOM error. 

% \revised{To better understand how the system design differences between \tool{} and GPUJoin impact runtime, we measured the most time-consuming components of both engines, as shown in Figure~\ref{fig:join_detail}. In the figure, ``Join'' represents the time spent on performing the join operation and materializing the results in both tools. In \tool{}, ``Merge'' refers to the time required to merge the delta tuples into the full, while ``Deduplication'' represents the time taken to remove duplicate tuples during iterative computation.
% The results show that \tool{} has a faster join speed compared to GPUJoin, due to the more efficient join algorithm and the fast range queries discussed in the second paragraph of Section~\ref{sec:join-impl}. The merge phase running time for both tools is similar, although merging the more complex HISA data structure in \tool{} takes more time, GPUJoin’s fused delta population and merge process (as discussed in the third paragraph pf Section~\ref{sect:pop-delta}) results in merging undeduplicated tuples, which increases both the number of tuples and the time required for merging.
% This fused design in GPUJoin causes deduplication to become a bottleneck, as it requires scanning the entire full relation to remove duplicates—an expensive operation when the tuple size is large. In contrast, \tool{}’s effective iterative evaluation design ensures that the deduplication time remains reasonable.}

\subsection{Context-Sensitive Program Analysis}
% comparing CSPA with souffle

Program analysis is one of the most popular applications of Datalog, with several state-of-the-art tools (DOOP~\cite{bravenboer2009doop}, cclyzer~\cite{balatsouras2016cclyzer}, and ddisasm~\cite{flores2020ddisasm}) implemented using Souffl\'e Datalog. To evaluate \tool{}'s potential application to program analysis, we implemented context-sensitive points-to analysis (\textit{CSPA}), reproducing the experiments of Graspan~\cite{Wang2017GraspanAS}. We elide the full query for space, it includes rules which initialize and propagate value flows, while simultaneously discovering alias information; Graspan uses method cloning (effecting inling) for context sensitivity.
%
% The \textit{CSPA} query, illustrated in Figure~\ref{fig:cspa-code}, includes two EDB relations: \textit{assign}, recording static variable assignments in the target program, and \textit{deref}, tracking pointer dereferences; these relations are constructed by a preprocessing pass. The rules of the query forms an Andersen-style data flow analysis, encapsulated the results in the IDB relation \textit{ValueFlow}, which delineates possible values of variables within specific calling contexts (encoded in the data via method cloning). The other two IDB relations, \textit{MemAlias} and \textit{ValueAlias}, record value and memory equivalences, respectively. The bottom five rules are used to initialize values in IDB relations; the mutually-recursive five upper rules declare the tuple population in the IDB rules. For instance, the first rule establishes the transitive nature of \textit{ValueFlow}, while the second rule addresses the scenario where two values, flowing to the identical variable under the same context, are deemed equivalent (a situation indicative of value flow conflation and precision loss in context-sensitive point-to analysis).
%
%
Using Graspan's data, we ran \textit{CSPA} on \textsf{httpd}, 
(a statically-linked subset of) \textsf{linux}, and \textsf{postgres}~\cite{Wang2017GraspanAS}. The results of our evaluation are presented in Table~\ref{tab:cspa_compare}; the table lists the size of input and output relations (second and third columns) and total running time (fourth column). All relation sizes match that of Souff\'e's.
% The large tuple sizes in both input and output make the query highly data-intensive, representing an attractive candidate for SIMD-based parallelization.

\yihao{We see a roughly 35-45$\times$ speedup versus Souffl\'e under ideal conditions (Souffl\'e use $32$ cores CPU with all optimizations, \tool{} running on NVIDIA H100 with maximum buffer size), demonstrating the significant performance advantage of \tool{} over the CPU-based Souffl\'e. This is mainly because the memory-heavy nature of program analysis workloads. Memory bandwidth and throughput significantly affect performance, and CPU systems usually have lower memory bandwidth. For example, the AMD Milan's bandwidth is around 190 GB/s, while a data center GPU like the NVIDIA H100 can have up to 3.35 TB/s. The lower memory speed and throughput prevent CPU-based systems from fully utilizing their high single-core performance. Our detailed analysis on CPU utilization further proves this, despite utilizing all 32 CPU cores during computation, the efficiency of Souffl\'e was relatively low, with CPU ratios of 571\%, 449\%, and 683\% for the \textsf{httpd}, \textsf{Linux}, and \textsf{PostgreSQL} datasets, respectively. Note that a 100\% utilization across 32 cores should yield a CPU ratio of 3200\%. }

Figure \ref{fig:join_detail} breaks down different phases of \tool{} when running our queries. In this figure, ``Deduplication'' shows time spent coalescing previously-discovered facts. ``Indexing Full/Delta'' represents the time required for creating hash indexing within \tool's hybrid data structure. `
`Merge Delta/Full'' captures the time spent on inserting all tuples from the delta relation into the full relation using GPU merge. 
Finally, ``Join'' represents the actual time spent performing join operations across all relations. 
\revised{The join (39\% of total runtime) and merge (42\% of total runtime) operations stand out as the most time-consuming phases in our pipeline, as depicted in the red and blue bar in the Figure~\ref{fig:join_detail}. Since performing joins between relations is computationally intensive and one of our main focuses, we expect this phase to occupy a significant portion of the total runtime. By leveraging the HISA datastructure, \tool{} is able to make these joins more efficient than previous GPU-based systems~\cite{shovon2023towards,shovon2022accelerating}. However, as the figure indicates, the merge operation is even more time-consuming than the join phase. 
This is due to the reason that lock-free style parallel path merge aglorithm we used require pre-allocating a result buffer that matches the combined size as total size of delta and full. Since the full relation in Datalog is typically large, this result buffer can be substantial. Allocating and freeing such a large buffer in every iteration is expensive, and, according to our detailed timing results, this step becomes even more time-consuming than the join and indexing operations in the query.
To partially address this challenge, we introduce an amortized buffer management scheme (described in Section~\ref{sec:impl-buffer}). While this approach doesn’t fully resolve the issue, it mitigates the need to allocate buffers at every iteration, thereby reducing the overall computational load.}

\subsection{\tool{} Across Different Hardware}
\label{sec:protablility}
To increase the performance portability of \tool{}, we translated it to a Heterogeneous-Compute Interface for Portability (HIP)~\cite{AMD_2024} based engine (\tool{}-HIP hereafter) with an identical API to our vanilla CUDA variant. This interface enables seamlessly switching \tool{} implementations from NVIDIA to HIP kernels. %In both CUDA and HIP, memory allocation plays a crucial role in enabling efficient parallel processing on GPUs. CUDA, developed by NVIDIA, and HIP, an open-source alternative primarily supported by AMD, share similarities in their approach to memory management but also exhibit notable differences. CUDA utilizes a specific set of memory allocation functions to allocate and deallocate memory on the GPU device, enabling data transfers between the CPU and GPU. CUDA's memory allocation is tightly integrated with its programming model, offering developers a streamlined experience when working with NVIDIA GPUs.
 The HIP version of \tool{}, while not matching the CUDA version's performance (due to missing libraries such as RMM \cite{librmm}), offers broad compatibility with different GPU architectures. This enables rapid deployment on the exascale systems, for example, Frontier and Aurora \cite{alcf-aurora,frontier}, which do not support CUDA.

\begin{table}[]
\fontsize{9}{11}\selectfont
\caption{\fontsize{9}{11}\selectfont Comparison of \tool{} and \tool{}-HIP running times across various GPU vendors and models } 
\begin{tabular}{@{}c|c|c|c||c|c@{}}
 % \multicolumn{2}{c}{}         & \multicolumn{4}{c}{Time (s)} \\
%\toprule
 \multicolumn{2}{c}{}          & \multicolumn{2}{c||}{\tool{} (NVIDIA)}    & \multicolumn{2}{c}{\tool{} (AMD)}      \\ 
%\midrule
 Query            & Dataset               & H100     & A100    & MI250    & MI50      \\ 
 \midrule
             & fe\_body       & 5.05     & 8.61    & 19.57    & 41.99     \\
SG           & BrightKite & 3.42     & 6.79    & 14.00    & 30.05     \\
             & fe\_sphere     & 2.36     & 4.64    & 8.48     & 19.426    \\ \midrule
             & \textsf{httpd}          & 1.33     & 2.73    & 6.75     & 15.27         \\
CSPA         & \textsf{linux}          & 0.39     & 0.77    & 1.39     & 3.32         \\
             & \textsf{postgres}     & 1.27     & 2.68    & 6.79     & 14.55         \\ \bottomrule
\end{tabular}
\label{tab:different-gpu}
\end{table}

To compare \tool{}'s performance across a range of data center GPUs, we ran \textit{SG} and \textit{CSPA} using \tool{} on the NVIDIA H100 and A100 and the AMD MI250 and MI50. The outcomes of these tests are presented in Table~\ref{tab:different-gpu}. The NVIDIA H100, successor to the A100, features more SM units ($114$ vs. $108$), double the FP32 cores per SM ($128$ vs. $64$), and improved memory bandwidth (2TB/s vs. 1.5TB/s). Columns 4 and 5 of our results show \tool{} benefits from these enhancements. % In every test, the H100 consistently delivers performance that is approximately twice as fast as the A100. Our results illustrate that \tool{} effectively harnesses the additional throughput offered by the latest GPUs.

The last two columns illustrate \tool{}'s performance on AMD GPUs. The MI250, with 104 computational units ($64$ core per unit) akin to NVIDIA’s SMs, shows comparable performance to the NVIDIA A100 in \textit{CSPA} tests but lags in \textit{SG} tests requiring more memory, performing at about half the A100's speed. The MI50, with half the capacity of the MI250, displays nearly half its performance, pointing towards the scalability of the approach. MI250 has similar total computational resources, it should have similar performance to the A100. However, the most important reason for the lower performance of the AMD MI250 is its dual chiplet design; since \tool{} is a single-GPU system, only half of the compute resources on the MI250 can be utilized, resulting in half the performance of the A100. Another reason is our use of the NVIDIA-specific RMM library in our implementation. Since AMD's ROCm does not support this library, we rely on manual memory pooling instead; we leave optimizing memory allocation in our HIP backend to future work.
 %In conclusion, our experiments demonstrate the robustness and scalability of \tool{} across diverse GPU architectures, including its capability to leverage advancements in GPU technology,

\begin{table}[]
    \centering
    \fontsize{9}{11}\selectfont
    \caption{\fontsize{9}{11}\selectfont Comparing Time Consuming Operation of \tool{} on AMD EPYC 7543P 32-core Zen 3 CPU and NVIDIA A100 GPU. Each running time in table are in seconds.}
    \label{tab:gdlog-cpu}
    \begin{tabular}{c|c c|c c|c c}
    \toprule
    & \multicolumn{2}{c}{Sort} & \multicolumn{2}{c}{Merge} & \multicolumn{2}{c}{Memory} \\
    \# Tuples  & \textbf{A100} & \textbf{Zen3} & \textbf{A100} & \textbf{Zen3} & \textbf{A100} & \textbf{Zen3} \\
    \hline
    1,000,000 & 0.12 & 1.09 & 0.03 & 0.06 & 0.03 & 0.02 \\
    10,000,000 & 0.39 & 7.5 & 0.08 & 0.64 & 0.17 & 0.05 \\
    50,000,000 & 1.63 & 30.09 & 0.18 & 1.96 & 0.11 & 0.88 \\
    100,000,000 & 2.9 & 64.02 & 0.3 & 3.56 & 0.18 & 1.7 \\
    500,000,000 & 15.66 & 351.4 & 1.21 & 15.68 & 0.82 & 8.59 \\
    \bottomrule
    \end{tabular}
\end{table}

Although fully implementing \tool{} on a CPU via SIMD is possible, we believe such an implementation would not achieve similar performance due to the significant memory bandwidth differences between the two hardware architectures. Due to the extensive potential engineering work, we did not implement a full CPU version of \tool{}. However, we did implement two \tool{}'s most computationally-expensive phases: sort, merge and implemented them using the latest version of Intel's oneTBB SIMD library (which uses CPU-specific vectorization)~\cite{onetbb}.

We ran both CPU and GPU implementations of sort and merge on different sizes of randomly generated tuples with two arities, collecting the total time of 100 runs. The results are presented in Table~\ref{tab:gdlog-cpu}. Overall, in all test cases, the GPU version of both sort and merge functions is around 10x to 20x faster than the CPU version. We believe this is primarily due to the significantly better memory bandwitdh offered by the GPU (1.5 TB/s vs. 0.19 TB/s). To investigate this, we collected the buffer allocation and initialization times for these two operations and listed them in the last column, which shows the memory allocation and initialization of the buffer to hold tuples used in the experiment. For all input sizes, the A100 GPU is consistently 10x faster than the EPYC 7543. We note that the performance increases mirror the memory bandwidth differences between the CPU and GPU; indeed, because modern Datalog applications (such as \textit{CSPA}) are memory-bound, we conclude GPUs offer an attractive implementation platform for modern Datalog engines.

\section{Conclusion and Future Work}

%Modern in-memory Datalog engines typically hit scalability walls around $8$--$16$ threads, due to limited memory bandwidth and the overhead of locking. The extreme parallel throughput offered by modern GPUs make them an attractive candidate for the implementation of high-performance backends for Datalog that modern power program analyses and similar applications. However, current approaches to GPU Datalog are stymied by incomplete implementations, limited expressivity, and a lack of data structures specialized to the concerns of modern Datalog evaluation (iterated joins and semi-na\"ive evaluation). 

We presented \tool{}, the first GPU-based Datalog engine able to achieve net-positive performance compared to SOTA Datalog implementations. Our library \tool{} (implemented both for CUDA and HIP) enables modern Datalog implementation techniques, namely compilation to relational algebra, range-indexed iterated joins, and optimized $n$-way joins. \tool{} is powered by the hash-indexed sorted array (HISA), a novel data structure we developed to support high-performance Datalog engines on the GPU. 
%HISA combines the algorithmic necessities of modern Datalogs while effectively leveraging the massive SIMD parallelism available to modern GPUs. 
We then used HISA to build \tool{} by employing two novel design points: temporarily materialized $n$-way joins and eager buffer management. We have evaluated \tool{} extensively compared to both CPU and GPU-based state-of-the-art; our experimental results demonstrate improvements of up to $45\times$ compared to an optimally-tuned CPU-based engine on context-sensitive points-to analysis of \textsf{PostgreSQL}.

% We plan several threads of future work. 
\revised{We notice recent work on running relational algebra and Datalog on multi-node systems, scaling them to thousands of nodes~\cite{sun2023communication, gilray2021compiling, kumar2020load, kumar2019distributed} using fine-tuned MPI all-to-all communication~\cite{fan2024configurable, fan2022optimizing}.
Following this trend, one of our future projects is multi-node, multi-GPU programming coordinated via MPI. Such an effort necessitates a heterogeneous decomposition of work across nodes while minimizing communication. We plan to address this by leveraging the coarse-grained (task-level) parallelism inherent in specific applications.} Additionally, we intend to extend \tool{} to support monotonic aggregation and implement recent join algorithms, such as free join~\cite{freejoin}.

%We believe our work shows the potential for GPUs to bring orders-of-magnitude performance improvements to analytics applications in program analysis, graph mining, medical reasoning, and similar deduction-heavy domains; we are currently applying \tool{} to large applications in several of these areas. As future work, we hope to explore multi-GPU joins, especially in the context of modern GPU interconnects and heterogeneous communication. We also hope to extend \tool{} to enable incremental computations over large data streams on modern data center GPUs.

\section{Acknowledgement}
This work was funded in part by NSF RII Track-4 award 2132013, NSF PPoSS planning award 2217036, NSF PPoSS large award 2316157 and, NSF collaborative research award 2221811. We are thankful to the ALCF’s Director’s Discretionary (DD) program for providing us with compute hours to run our experiments on the Polaris supercomputer and the Joint Laboratory for System Evaluation (JLSE) located at the Argonne National Laboratory. 
% This material is based upon work supported by the Defense Advanced Research Projects Agency (DARPA) under Contract No. N66001-21-C-4023. Any opinions, findings and conclusions or recommendations expressed in this material are those of the author(s) and do not necessarily reflect the views of DARPA.

\bibliographystyle{plain}
\bibliography{main}

\end{document}